%% file: main.tex
\documentclass[numberedappendix,tighten,iop]{openjournal}

\usepackage[T1]{fontenc}
\usepackage{lmodern}
\usepackage{microtype}
\usepackage{textcomp}

\usepackage{savesym}
\savesymbol{tablenum}
\usepackage{siunitx}
\restoresymbol{SIX}{tablenum}
\DeclareSIUnit\parsec{pc}
\usepackage{xcolor}
\usepackage{amsmath}
\usepackage{bigdelim}
\usepackage{booktabs}
\usepackage{hyperref}
\hypersetup{colorlinks=true,linkcolor=blue,citecolor=blue,filecolor=blue,urlcolor=blue}
\usepackage[capitalise]{cleveref}
\usepackage{natbib}

\include{macros}
\include{journal_macros}

\begin{document}
\title{Statistical Uncertainties of the N\textsubscript{DW} = 1 QCD Axion Mass Window from Topological Defects}

\author{Sebastian Hoof\textsuperscript{1}}
\author{Jana Riess\textsuperscript{1}}
\author{David J.~E. Marsh\textsuperscript{2,1}}

\affiliation{\textsuperscript{1}Institut f\"{u}r Astrophysik,  Georg-August-Universit\"{a}t {G\"{o}ttingen}, Friedrich-Hund-Platz~1, 37077\ {G\"{o}ttingen}, Germany}
\affiliation{\textsuperscript{2}Department of Physics, King's College London, Strand, London WC2R 2LS, United Kingdom}
\thanks{Emails: \aem{hoof@uni-goettingen.de}, \aem{jriess@astro.physik.uni-goettingen.de}, \aem{david.j.marsh@kcl.ac.uk}}

\begin{abstract}
We review results from QCD axion string and domain wall simulations and propagate the associated uncertainties, including QCD uncertainties, into the calculation of the axion relic density. We compare two different sets of studies and, using cosmological constraints, perform statistical inference on the axion mass window in the post-inflationary Peccei--Quinn symmetry breaking scenario. For dark matter axions in recent simulations inferring a moderately infrared-dominated spectrum, this leads to a median dark matter axion mass of 0.50\,meV, while the 95\% credible interval at highest posterior density is between 0.48 and 0.52\,meV. For alternative simulations including in addition string-domain wall decays (but with different overall inference on the spectrum), these numbers are 0.22\,meV and [0.16, 0.27]\,meV. Relaxing the condition that axions are all of the dark matter, the axion mass window is completed by an upper bound of around 80\,meV, which comes from dark radiation constraints. This confirms that the axion mass can be constrained rather precisely regarding statistical uncertainties and further calls for a more detailed analysis of the various sources of systematic uncertainties plaguing the simulations.
\end{abstract}

\maketitle
%\flushbottom

\section{Introduction}\label{sec:intro}
QCD axions~\citep{PQ1,PQ2,Weinberg:1977ma,Wilczek:1977pj} are a well-known solution to the Strong CP problem of the Standard Model~(SM) and excellent dark matter (DM) candidates~\citep{Preskill:1982cy,Abbott:1982af,Dine:1982ah,Turner:1983he,Turner:1985si}. If the associated Peccei--Quinn~(PQ) symmetry breaks after the end of inflation, and is never restored, the value of the axion relic density can in principle be calculated from a single unknown model parameter: the axion decay constant~$f_a$~(assuming the PQ~field quartic coupling to be $\lr \sim 1$). In this scenario, the total axion relic density consists of axions from the so-called realignment mechanism and from topologically nontrivial field configurations, known as cosmic strings and domain walls~(DWs). Such defects form dynamically during the PQ phase transition~\citep{Kibble:1976sj} and emit axions during their subsequent evolution~\citep{Davis:1985pt,Davis:1986xc,Harari:1987ht,astro-ph/9311017}.

Accurately predicting the axion relic density in the post-inflation PQ~breaking scenario allows to set a lower limit on the axion mass from the observed amount of DM in the Universe, or even to predict the axion mass assuming it is all the DM.
There has been a debate on the behavior of the string network and, consequently, on the significance of the string contribution to the axion relic density for almost three decades.
A~complete numerical simulation of the strings to the physically relevant regime is currently not possible due of the large separation of scales in the problem of order~$\ee^{70} \sim 10^{30}$, set by the ratio of the size of the string core (near the PQ scale) to the Hubble scale when the axion field becomes dynamical (near the axion mass scale).
State-of-the-art simulations can only access separations of up to about $\ee^{8} \approx 3000$~\citep{2007.04990}, and one therefore has to rely on extrapolations over many orders of magnitude.
While the extrapolation will introduce sizeable systematic uncertainties, it is not a hopeless endeavor due to the conjectured existence of an attractor solution for the string scaling \citep[e.g.][]{Kibble:1976sj,Kibble:1980mv,Vilenkin:1981kz}.

A crucial quantity in this extrapolation is the shape of the instantaneous axion emission spectrum, which is still a subject of debate~\citep{Davis:1985pt,Davis:1986xc,Harari:1987ht,astro-ph/9311017,astro-ph/9403018,hep-ph/9807428,0812.1929,1012.5502,1707.05566,1708.07521,1806.04677,1809.09241,1906.00967,1908.03522,2007.04990,2102.07723,2108.05368}.
Effectively, the shape of the power spectrum can be characterized by its power law index $q$ for the relevant wave numbers $k$, where the power spectrum has a $1/k^q$ behavior.   
Some simulation studies found that the spectrum is UV-dominated, i.e.\ by larger energies, leading to a sub-dominant contribution to the relic density from strings, expressed by $q < 1$~\citep[e.g.][]{1707.05566,1708.07521,1906.00967}.\footnote{While this work was in preparation, new results by \cite{2108.05368} appeared, indicating indicating $q \sim 1$  and superseding \cite{1906.00967}. We therefore do not include the results of \cite{1906.00967} here.}
However, other authors observed an IR-dominated spectrum with $q \gtrsim 1$ in their simulations, which would cause the contribution from topological defects to be similar or dominant compared to the realignment mechanism~\cite[e.g.][]{1012.5502,2007.04990,2108.05368}.

So far, discussion has mostly focused on qualitative aspects of the behavior of the strings, the existence of a scaling solution, and the shape of the spectrum.
While these also rely on the quantitative nature of the related parameters, no extensive analysis of their statistical and systematic uncertainties has been put forward.

The goal of the present work is to consider numerical results from simulations of cosmic strings and DWs in a common framework, and to compute the energy density in axions from realignment and topological defects, propagate the relevant statistical uncertainties and perform a likelihood analysis of cosmological constraints.
In this sense, our work is in the spirit of e.g.\ \cite{1202.5851,1412.0789} but extended by the formalism and findings of \cite{1806.04677,2007.04990}.
This allows us to identify a minimal set of parameters for characterizing e.g.\ the string spectrum and for considering the various sources of measurable uncertainties.
Doing so allows us to estimate the axion mass window as informed by the amount of DM and the number of relativistic species in the Universe from \textit{Planck}'s measurement of the cosmic microwave background~(CMB) anisotropies~\citep{1807.06209}.

In Sec.~\ref{sec:qcd_axion} we outline the relevant properties of the QCD axion. In Sec.~\ref{sec:equations}, we summarize the equations for computing the realignment and topological defect contributions to the axion energy density today. Sections~\ref{sec:params} and~\ref{sec:constraints} provide details on extracting the parameters estimates and constraints. Section~\ref{sec:results} is to present and discuss our findings before we conclude by commenting on our results and their relevance with respect to systematics in Sec.~\ref{sec:conclusions}.

\section{QCD axion properties}\label{sec:qcd_axion}
For our considerations, we only need to know a few of the calculated QCD~axion properties. One is the QCD axion mass, which (at zero temperature) can be derived from chiral perturbation theory~(ChPT), following the Gell-Mann--Oakes--Renner relation~\citep{PhysRev.175.2195}:
\begin{equation}
    f_a^2 m_{a,0}^2 = \chi_0 \, , \label{eq:topo_suscept}
\end{equation}
where $\chi_0$ is the zero-temperature QCD topological susceptibility, $f_a = v_\text{PQ}/\NDW$ is the axion decay constant, $v_\text{PQ}$ is the PQ field vacuum expectation value, and $\NDW = 2N$ is the domain wall number, given by twice the color anomaly coefficient~$N$.\footnote{In what follows we assume $\NDW = 1$, and consider only strings and unstable DWs. With $\NDW > 1$, DWs are stable and an arbitrary biasing potential must be introduced such that DWs do not produce more DM than the observed amount. This additional freedom makes a statistical analysis of the case $\NDW > 1$ redundant. The axion mass prediction in this scenario is controlled by the choice of biasing, and a fine tuning measure for the neutron electric dipole moment \citep{1207.3166,1412.0789}.}

\cite{Weinberg:1977ma} first used $\chi_0$ to estimate the axion mass, and the calculation has been refined over the years~\cite[e.g.][]{1511.02867,1812.01008}. \cite{1812.01008} find that
\begin{align}
    \chi_0 &= \chi_0^\text{\tiny LO} (1 + \delta^\text{\tiny NLO} + \delta^\text{\tiny NNLO} + \delta^e) \, , \label{eq:qcdaxion:zeromass:nnlo} \\
    m_{a,0} &= \frac{\sqrt{\chi_0}}{f_a} = \SI{5.69(5)}{\micro\eV} \left( \frac{\SI{e12}{\GeV}}{f_a} \right) \, , \label{eq:qcdaxion:zeromass}
\end{align}
where the corrections in Eq.~\eqref{eq:qcdaxion:zeromass:nnlo} are given in \cite{1812.01008} and lead to the quoted uncertainty in Eq.~\eqref{eq:qcdaxion:zeromass}, which we discuss further in Sec.~\ref{sec:paramsqcd}.

The QCD~axion mass exhibits a temperature dependence, which can be parameterized as
\begin{equation}
    m_a(T) = \frac{\sqrt{\chi(T)}}{f_a} = \frac{\sqrt{\chi_0}}{f_a} \, 
	\begin{cases}
	\hfil 1 & \text{if } T \leq T_\chi \\
		\left(\frac{T_\chi}{T}\right)^{p/2} & \text{otherwise}
	\end{cases} \, . \label{eq:qcdaxion:mass}
\end{equation}
From the dilute instanton gas approximation~\citep{1981_Gross}, we know that $p \sim 8$~\citep{Preskill:1982cy,Abbott:1982af,Dine:1982ah,Turner:1985si}, and $T_\chi$ should the of the same order as e.g.\ the QCD crossover temperature, which is $T_\text{QCD,c} \approx \SI{157}{\MeV}$~\citep{1812.08235,2002.02821}. In the following we fit the temperature dependence of $\chi(T)$ to the lattice QCD results of \cite{LatticQCD4Cosmo}, as described in Sec.~\ref{sec:paramsqcd}.

QCD~axions can have interactions with several other particles. While the most useful coupling for axion searches is that to photons, we only need the axion coupling to gluons and pions. The gluon coupling is given by
\begin{equation}
    \mathcal{L} \supset - \frac{\alphaS}{8\pi f_a} \, \phi \,  G_{\mu\nu}\widetilde{G}^{\mu\nu} \, ,
\end{equation}
where $\phi$ is the axion field, $G_{\mu\nu}$ is the gluon field strength tensor and $\widetilde{G}^{\mu\nu}$ its dual, and $\alphaS$ is the strong-force fine-structure constant. In turn, we define the model-independent part of the axion-pion coupling in the effective Lagrangian as~\citep{2003.01100}
\begin{equation}
    \mathcal{L}_\text{eff} \supset \frac{\Capi}{f_a f_\pi} \partial_\mu a [ 2 (\partial^\mu \pi^0) \pi^+ \pi^- - \pi^0 (\partial^\mu \pi^+) \pi^- - \pi^0 \pi^+ (\partial^\mu \pi^-) ]
\end{equation}
with the coupling constant
\begin{equation}
    \gapi \equiv \Capi \, \frac{m_\pi}{f_a} = \frac{1}{3} \left(\frac{1-z}{1+z}\right) \frac{m_\pi}{f_a}\, ,
\end{equation}
where $m_\pi$ is the neutral pion mass, and $z=m_u/m_d$ is the ratio of the up and down quark masses. We do not consider additional, model-dependent terms that would arise in non-hadronic axion models~[see e.g.\ \cite{2003.01100} for details].

\section{The QCD axion relic density}\label{sec:equations}
\begin{figure*}
	\centering
	\includegraphics[width=0.9\textwidth]{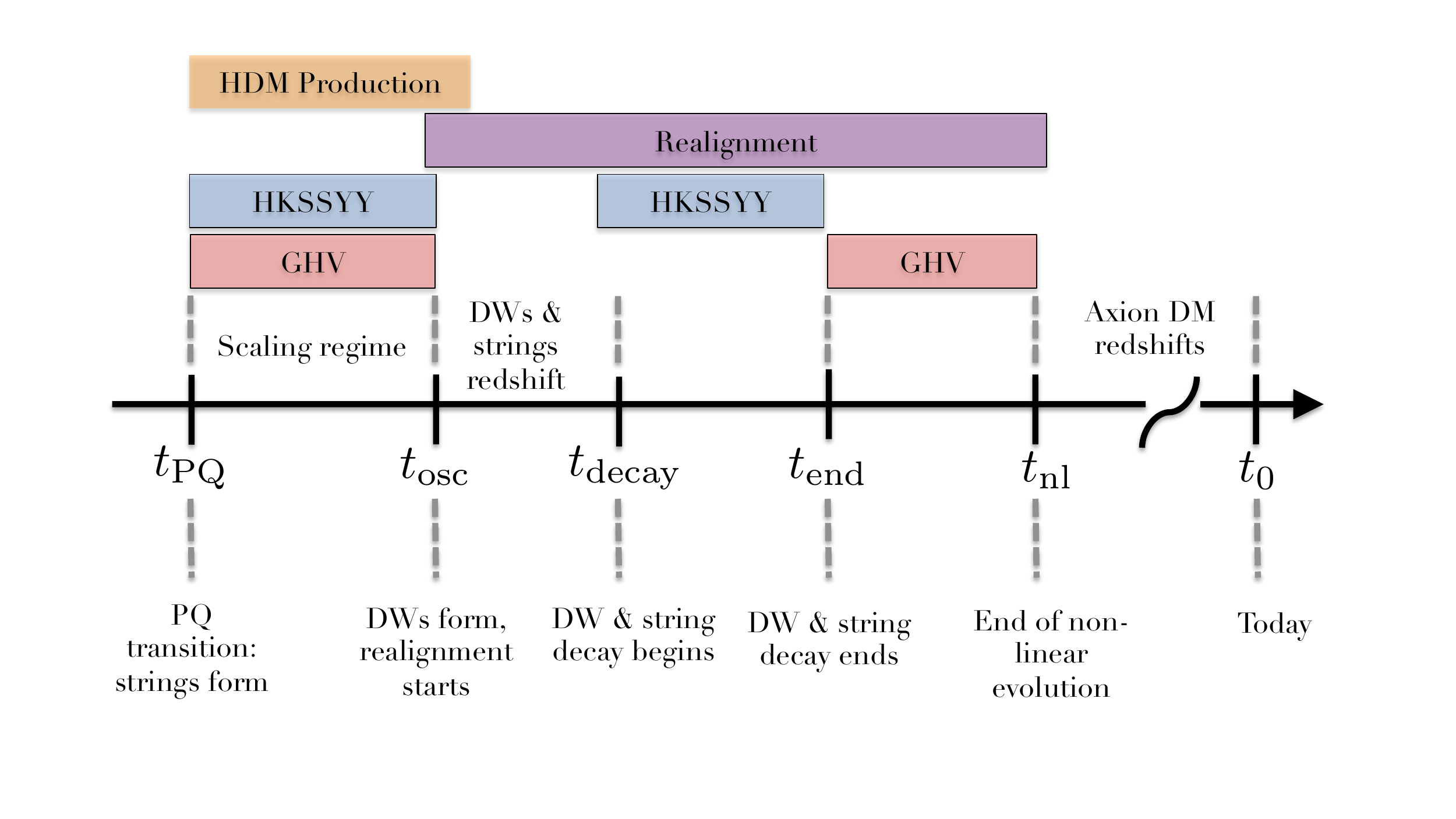}
	\caption{Schematic timeline of axion production in the post-inflation Peccei--Quinn symmetry breaking scenario. The labels HDM, GHV, and HKSSYY refer to hot dark matter, \cite{1806.04677,2007.04990}, and \cite{1012.5502,1202.5851,1412.0789}, respectively. More detailed explanations for the different time windows can be found in the main text.\label{fig:flowchart}}
\end{figure*}
Here, we provide the equations used in this work to calculate the different contributions to the axion energy density. They form the basis of our framework for re-casting and comparing the findings of different works on the string spectrum. A graphical overview of the different mechanisms that we consider, and the associated time scales, is presented in \cref{fig:flowchart}.

The axion energy density today, $\rho_a(T_0)$, at temperature~$T_0 = \SIsmp{2.7255(6)}{\K}$~\citep{0911.1955} is usually expressed relative to the critical density $\rhoc = 3\mpl^2H_0^2$, i.e.\ as $\Omega_a = \rho_a/\rhoc$, where $H_0 = 100h\,\si{\km\,\s^{-1} \mega\parsec^{-1}}$ with the relative Hubble constant~$h$ and $\mpl \approx \SI{2.435e18}{\GeV}$ is the reduced Planck mass. Under the assumption of (approximate) entropy conservation, the comoving number density of axions is conserved, and we can use this fact to scale the axion number densities between physical times, or in our case also temperatures~$T$ and $T_0$:
\begin{equation}
    \rho_a (T_0) = \rho_a(T) \, \frac{m_{a,0}}{m_{a}(T)} \, \frac{\gS(T_0)}{\gS(T)} \, \left(\frac{T_0}{T}\right)^3 \, , \label{eq:edescaling}
\end{equation}
where $\gS$ is the number of effective relativistic degrees of freedom for the entropy density. In what follows, we will quote the axion energy densities and temperatures from where the scaling to today's energy density is possible. Note that we ignore the small uncertainty on~$T_0$ in what follows and use the temperature-dependent values of $\gR$ (the number of effective relativistic degrees of freedom) and $\gS$ from \cite{LatticQCD4Cosmo}~(also without uncertainties; see Appendix~\ref{sec:g_star_appendix} for details on our implementation).

\subsection{Realignment}\label{sec:realignment}
The realignment contribution to the relic density comes from the zero-mode of the axion field~$\phi$. Its evolution is given by the Klein--Gordon equation on a 
Friedmann--{Lema\^itre}--Robertson--Walker background
\begin{equation}
    \ddot{\phi} + 3 H \dot{\phi} + V'(\phi) = 0 \, , \label{eq:klein_gordon}
\end{equation}
where $V'(\phi)$ is the derivative of the effective axion potential~$V$ with respect to the axion field~$\phi$. We use the simple one-instanton potential for \cref{eq:klein_gordon},
\begin{equation}
    V(\phi) = f_a^2 m_a^2(T) \left[1 - \cos(\phi / f_a)\right] \, , \label{eq:kge:potential}
\end{equation}
where $m_a(T)$ is the temperature-dependent axion mass from \cref{eq:qcdaxion:mass}.
The time evolution of the temperature during radiation domination can be obtained from the Friedmann equation
\begin{align}
    \label{eq:friedmann}
    3 \mpl^2 H^2 = \rho_\text{rad} = \frac{\pi^2}{30} \, \gR(T) \, T^4 \, ,
\end{align}
where $\gR$ are the effective relativistic degrees of freedom for the energy density, and the usual initial conditions $\phi(t = 0) = f_a \thetai$ and $\dot{\phi}(t = 0) = 0$~\citep{book_kolbturner,book_weinberg_cosmo}, where $\thetai \in (-\pi,\pi]$ is the so-called misalignment angle.

At early times, the Hubble friction in Eq.~\eqref{eq:klein_gordon} dominates, and the field remains constant. At some time $\tosc$, approximately defined via
\begin{align}
   \label{eq:osc_cond} 
   3 H(\tosc) = m_a(\Tosc) \, ,
\end{align}
where $\Tosc$ is the corresponding temperature, the axion field starts to oscillate. At later times~$t \gtrsim \tosc$~(and temperatures~$T \lesssim \Tosc$), when $m_a(T) \gg H(T)$, the comoving number of axions is conserved and, under the assumption of entropy conservation, we can scale the energy density in axions,
\begin{align}
    \rho_a^\text{re} = \frac{1}{2}\dot{\phi}^2 + V(\phi) \, ,
\end{align}
to its value today via Eq.~\eqref{eq:edescaling}. Details on how we solve Eq.~\eqref{eq:klein_gordon} numerically can be found in Appendix~\ref{sec:klein_gordon_appendix}.

In the post-inflation PQ~symmetry breaking scenario that we consider, the axion field takes on random values in the large number of causally-disconnected patches in the early Universe~\citep{Turner:1985si}. If the potential is initially close to zero, and switched on sufficiently fast, $\thetai$ follows a uniform distribution $\thetai \sim \mathcal{U}(-
\pi,\pi)$.

As a consequence, the axion energy density on cosmological scales today can be calculated as the average over the huge number of causally-disconnected patches. This average can be calculated conveniently via the integral
\begin{equation} 
    \langle \rho_a^\text{re} \rangle_{\thetai} \approx \frac{1}{2\pi} \int_{-\pi}^{\pi} \! \rho_a^\text{re}(f_a, \thetai; \sqrt{\chi_0}, T_\chi, p) \; \dd\thetai \, . \label{eq:rd:realign}
\end{equation}

\begin{figure}
    \centering
    \includegraphics[width=4in]{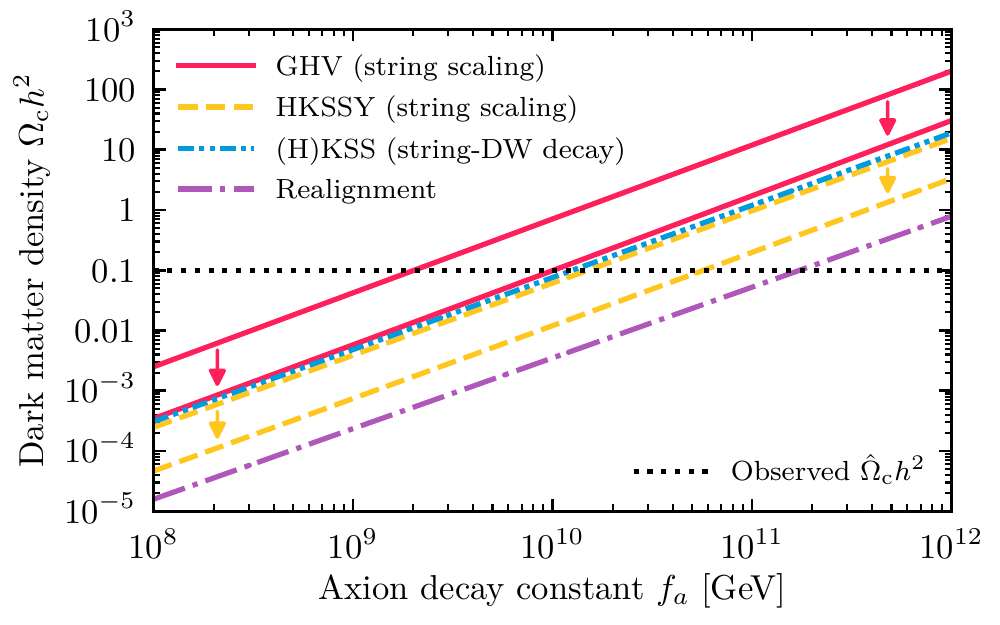}
    \caption{The different contributions to the axion relic density as functions of $f_a$ for typical values of the other parameters. We show realignment~(dashed-dotted purple), string-DW decay~(dashed-dotted blue), as well as string scaling contributions for the different spectra~(solid red and dashed yellow). For the latter, arrows indicate the reduction of the energy density by the the nonlinear transient. The dark matter density is shown as a horizontal, dotted black line.}
    \label{fig:axion_energy_density}
\end{figure}
We show the realignment energy density from Eq.~\eqref{eq:rd:realign} as a function of~$f_a$ in Fig.~\ref{fig:axion_energy_density}.

\subsection{Axion emission from string scaling}\label{sec:strings}
The axion field~$\phi$ arises as the phase of a complex scalar field~$\Phi$, whose potential for $\NDW = 1$ can be written as
\begin{align}
    V(\Phi) = \lr \left(\left|\Phi\right|^2 - \frac{f_a^2}{2}\right)^2 \, . \label{eq:sombrero}
\end{align}
Note that this potential respects the $\mathrm{U}(1)$ PQ~symmetry and that the quartic coupling $\lr$ is a free model parameter. We may re-write the potential in terms of the radial mass~$m_r$ via $\lr \equiv m_r^2/2f_a^2$. While it is usually assumed that $\lr \sim \order{1}$, it should be noted that this parameter may be orders of magnitude smaller. For example, the so-called SMASH model requires $\num{5e-13} < \lr < \num{5e-10}$~\citep{1610.01639}.

After PQ~symmetry breaking, we may expand~$\Phi$ around its vacuum expectation value~$|\langle\Phi\rangle|^2 = f_a^2/2$ such that
\begin{equation}
    \Phi(x) = \frac{r(x) + f_a}{\sqrt{2}} \; \ee^{i\phi(x)/f_a} \, , \label{eq:ssbfield}
\end{equation}
where $r$~is the radial mode and $\phi$~is the dynamical axion field~(as used before). 

The randomness of the axion field after symmetry breaking leads to the formation of cosmic strings~\citep{Kibble:1976sj,Kibble:1980mv,Vilenkin:1982ks}. The strings are characterized by their tension~(energy per unit length), which for isolated, infinitely long strings at late times can be estimated as~\citep{Vilenkin:1982ks}
\begin{equation}
    \label{eq:string_tension_theory}
    \mu \sim \pi f_a^2 \log \left( \frac{m_r}{H} \right) \, .
\end{equation}

Early simulations of cosmic strings~\cite[e.g.][]{1985PhRvL..54.1868A,1988PhRvL..60..257B} already suggested that strings enter a scaling regime, where the number of strings per Hubble volume stays constant. The existence of this scaling solution is a crucial ingredient for making the prediction of the axion energy density today independent from the system's initial conditions.

However, computing the exact evolution of and interactions between the strings at earlier times are rather involved tasks, such that numerical simulations of the full axion field equations are needed for a more precise prediction of e.g.\ the string tension compared to \cref{eq:string_tension_theory}.

For the remaining part of this section, we follow the approach and parameterization developed in \cite{1806.04677,2007.04990}~(hereafter also referred to as GHV-I and GHV-II, and collectively as GHV). The authors capture the non-trivial evolution of the string network by defining the average number of strings per Hubble patch~$\xi$ and the effective string tension as
\begin{align}
    \mu_\text{eff}(t) = \frac{\rho_\text{s}(t) \, t^2}{\xi(t)} \quad
    \text{with} \quad \xi(t) \equiv \lim_{V \rightarrow \infty} \left[ \frac{L_\text{tot}(V) \, t^2}{V} \right]\, , \label{eq:scaling_parameter}
\end{align}
where $L_\text{tot}(V)$ is the total length of all strings in volume~$V$ and $\rho_\text{s}$ is the energy density of strings. The (initial) time dependence of $\mu_\text{eff}$ can be determined from numerical simulations. GHV point out that a useful way to parameterize this dependence is, in analogy to \cref{eq:string_tension_theory},
\begin{equation}
    \mu_\text{eff}(t) = \gamma_1(t) \, \pi f_a^2 \log \left(\frac{\gamma_2(t) \, m_r}{H \sqrt{\xi(t)}}\right) \, ,
\end{equation}
where $\gamma_1(t)$ and $\gamma_2(t)$ are \textit{a~priori} unknown function and $\gamma_2(t)/\sqrt{\xi(t)}$ is supposed to capture the time dependence of~$\mu_\text{eff}$. The $\gamma_1$ factor accounts for the finite distance between the strings~(i.e.\ acts as an IR cutoff), the non-trivial shape of the strings, and possibly other effects. The $\gamma_2$ factor implements another IR cutoff, accounting for the average distance between straight uniformly distributed strings~(compared to a single string where the cutoff would be the size of the simulation volume).

The results of GHV-II imply that both $\gamma_1$ and $\gamma_2$ can be taken as constants, where we set $\gamma_1 = 1.3$ and $\gamma_2 = 1/\sqrt{4\pi}$ following appendix~E.1 of GHV-II. While GHV find that e.g.\ $\gamma_1 \sim \numrange[range-phrase=\text{--}]{1.3}{1.4}$, we expect that the systematic uncertainties associated with this choice of parameters will be re-absorbed through the definition of the energy density. This is because it is possible to estimate the energy density in strings directly via simulations. Doing so, the energy density in strings is trivially given by
\begin{equation}
    \rho_\text{s}(t) = \frac{\mu_\text{eff}(t) \, \xi(t)}{t^2} \, .
    \label{eq:energy_density_strings}
\end{equation}

The crucial point is now that maintaining the scaling solution requires that the strings radiate energy at a rate~$\Gamma$, which happens in the form of axions and radial modes. To leading order, the continuity equations for strings during the scaling regime is
\begin{equation}
    \dot{\rho}_\text{s} + 2 H \rho_\text{s} = -\Gamma \, .
\end{equation}

The rate $\Gamma$ can be found by comparing the derivatives of the energy density of the radiating string network to that of a ``free'' network. While the total emission rate is in principle given by the sum of emission rates into axions and radial modes, $\Gamma = \Gamma_a + \Gamma_r$, it turns out that the emission of axions is dominant i.e.\ $\Gamma \approx \Gamma_a$. It can be shown that~\citep{1806.04677}
\begin{equation}
    \Gamma_a \approx \Gamma  \simeq 2 H \rho_\text{s} = \frac{\mu_\text{eff}(t) \, \xi(t)}{t^3} \, , \label{eq:emission_rate}
\end{equation}
where the last equality is valid as long as the Universe is still radiation-dominated (since $H = 1/2t$ from solution of the Friedmann equation).
The number density of axions, at times when they are still effectively massless, can be calculated from
\begin{equation}
    \rhos(t) = \int \! \dd t^\prime \, \left(\frac{R(t^\prime)}{R(t)}\right)^4 \Gamma_a(t^\prime) \label{eq:energy_spectrum_rel} \, ,
\end{equation}
where $R$ is the cosmological scale factor. More generally, the rate in \cref{eq:emission_rate} should be understood to be derived from $\Gamma(t) = \int \! \dd k \, \del \Gamma (k,t)/\del k$ since one needs to take the axion momentum into account. It turns out that a useful way to parameterize this dependence is via~\citep{1806.04677}
\begin{equation}
    \frac{\del \Gamma_a(t,k)}{\del k} = \frac{\Gamma_a(t)}{H(t)} \, F\left[\frac{k}{H(t)},\frac{m_r}{H(t)}\right] \, , \label{eq:def_F}
\end{equation}
where the \emph{instantaneous emission spectrum}~$F$, with $\int \! \dd x \, F[x,m_r/H] = 1$, characterizes the spectrum, and $\Gamma_a(t)$ is given by \cref{eq:emission_rate}.
This allows us to calculate the axion number density from string scaling as 
\begin{align}
    n_a^\text{str} = \int \! \frac{\dd k}{k} \, \frac{\del \rhos}{\del k} = \int \! \dd t^\prime \, \left(\frac{R(t^\prime)}{R(t)}\right)^3 \frac{\Gamma_a(t^\prime)}{H(t^\prime)} \int \! \frac{\dd x}{x} \, F\left[x, \frac{m_r}{H(t^\prime)}\right] \, . \label{eq:number_density_from_spectrum}
\end{align}

Since the dynamics of the string network is governed by two energy scales only, namely the Hubble scale~$H$ and the string mass~$m_r$ (the inverse of the string core size), it is reasonable to assume that $F[x,y]$ is a power law in $x \equiv k/H$ between $x_0 < x \lesssim y \equiv m_r/H$ with some cutoff~$x_0$ that falls off outside of this peak region~\citep{1806.04677}:
\begin{align}
    \label{eq:instantaneous_emission_spectrum_form}
    F[x, y] = \frac{1}{x_0} \frac{q - 1}{1 - \left(x_0/y\right)^{q - 1}} \left( \frac{x_0}{x} \right)^q \, .
\end{align}
This leads to the result~\citep{1806.04677}
\begin{align}
    n_a^\text{str} \simeq \frac{8 H \,  \mu_\text{eff} \, \xi}{x_0} \frac{1 - q^{-1}}{1 - (2q - 1) \, \exp\left[(1 - q) \log\left(\frac{m_r}{x_0 H}\right)\right]}
    \label{eq:gorghettos_number_density_formula}
\end{align}
for large values of $\ell \equiv \log(m_r / H)$ such as the physically relevant scale around $\ell \sim 70$. 

\begin{figure*}
    \centering
    {
    \hfill
    \includegraphics[width=3in]{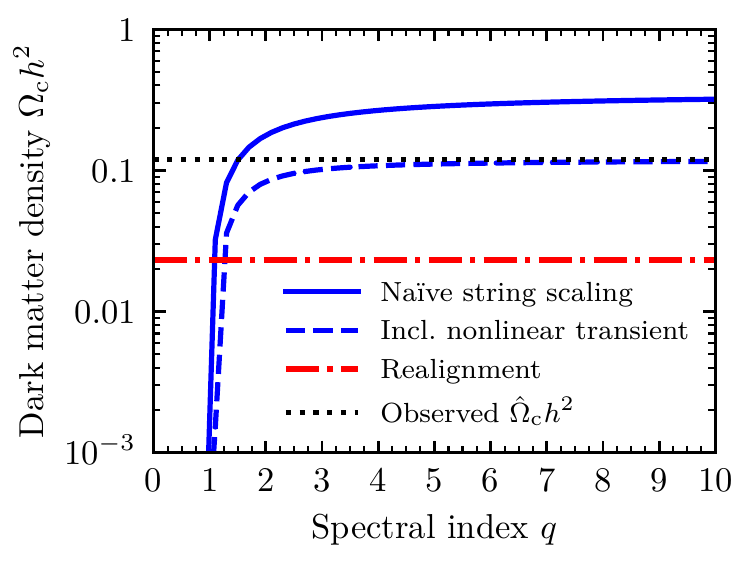}
    \hfill
    \includegraphics[width=3in]{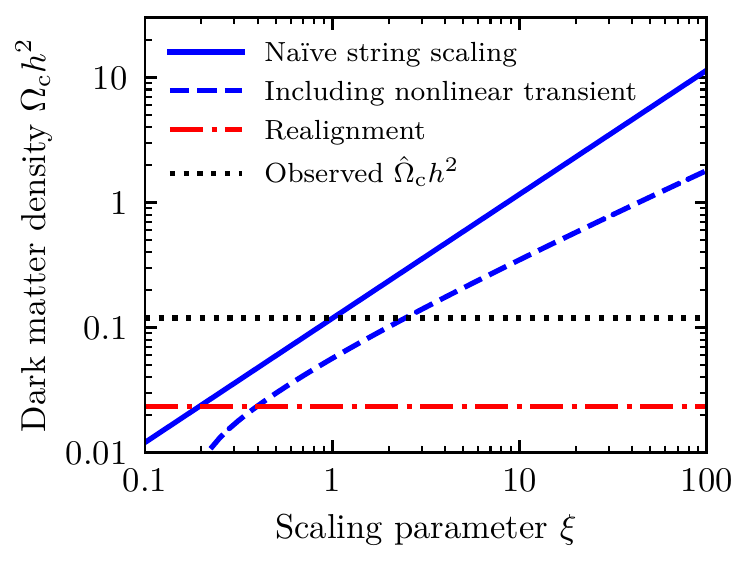}
    \hfill
    }
    \caption{The axion relic density as a function of the spectral index $q$ (\textit{left}; $\xi = 1$) and scaling parameter $\xi$ (\textit{right}; $q = 1.5$) for $f_a = \SI{5e10}{\GeV}$ and $x_0 = 10$. We show the result with~(dashed blue) and without~(solid blue) taking into account the nonlinear transient. We also indicate the observed dark matter density~(dotted black) and the corresponding realignment density~(dashed-dotted red).}
    \label{fig:string_relic_density_dependence}
\end{figure*}

In summary, one only need to extrapolate the values of the spectral cutoff~$x_0$, index~$q$, and the scaling parameter~$\xi$ from numerical simulations of the string network. The single most important parameter out of these is~$q$. For $q > 1$, the contribution from strings to the axion relic density tends to dominate over the realignment contribution, while for $q < 1$, the opposite is true~(cf.\ left panel of \cref{fig:string_relic_density_dependence}).

Since the number of axions from strings should be conserved after the decay of the strings, the relic density from strings today can be obtained from Eqs.~\eqref{eq:edescaling} and~\eqref{eq:gorghettos_number_density_formula}. The result is shown in \cref{fig:axion_energy_density}, and the dependence of $\Omega_a^\text{str}$ on $q$ and~$\xi$ is shown in \cref{fig:string_relic_density_dependence}.

\subsection{Nonlinear transient}
GHV-II point out that, at $t_*$ defined via $m_a(t_*) = H(t_*)$, the axion potential is still not relevant for the field evolution as long as the gradient term in the full axion field equation dominates.
This delay suppresses the comoving axion number density (when gradients dominate the energy density redshifts like radiation), while the axion mass, and hence the energy required to produce axions, grows in the meantime.
Ignoring the decaying strings and DWs, the spectrum is thus effectively redshifted as relativistic radiation, until some time $\tnl$, when the potential becomes effective, and non-linear evolution ends. This time is implicitly given by the condition~\citep{2007.04990}
\begin{equation}
    \int_0^{c_m m_a(\tnl)} \dd k \; \frac{\partial \rho_a(\tnl)}{\partial k}= c_V \, f_a^2 m_a^2(\tnl) \, ,
    \label{eq:non_linear_cond}
\end{equation}
where the numerical constants $c_m$ and $c_V$ can be determined from numerical simulations. Following \cite{2007.04990}, \cref{eq:non_linear_cond} can be further expanded, and we solve it numerically via root-finding for $m_a(\tnl)/H_*$, where the initial guess is given by
\begin{equation}
    \label{eq:non_linear_guess}
    \frac{m_a(\tnl)}{H_*} = \left[ \frac{W_{-1} \left(-\frac{c_V \left(1 + \frac{2}{p + 2}\right)}{4 \pi \xi_* \log_*} \left( \frac{x_0}{c_m} \right)^{2(1 + \frac{2}{p+2})} \right)}{- \frac{c_V \left(1 + \frac{2}{p + 2}\right)}{4 \pi \xi_* \log_*}} \right]^{\frac{1}{2} \frac{p}{p+4}} \, ,
\end{equation}
with $W_{-1}$ being the Lambert~$W$ function on branch~$-1$. The corrected number density at $\Tnl$ is then given by
\begin{equation}
    n_a^\text{str}(\Tnl) = c_n \, c_V \, f_a^2 \, m_a(\Tnl) \, , \label{eq:ede_a_str_nl}
\end{equation}
where $c_n$ is another numerical constant, which captures all remaining matching effects of the nonlinear transient. \Cref{eq:ede_a_str_nl} can thus be viewed as a correction to the axion energy density from strings, the effect of which can be seen in \cref{fig:axion_energy_density,fig:string_relic_density_dependence}.

\subsection{String-domain wall decays}\label{sec:dws}
Domain walls form when the axion mass switches on around the QCD~phase transition.
Once the DW~tension is of order the string tension, the unstable system of strings and DWs with $\NDW = 1$ starts decaying whilst emitting axions. The corresponding condition
\begin{equation}
    \sigmaDW(\tdecay) = \mu_\text{eff}(\tdecay) / \tdecay \, , \label{eq:decay_condition}
\end{equation}
implicitly defines the timescale~$\tdecay > \tosc$. The DW tension $\sigmaDW$ can found from~\citep{1511.02867}
\begin{align}
    \sigmaDW(t) = \mathcal{I} \, f_a^2 m_a(t) \quad
    \text{with} \quad \mathcal{I} &\equiv \frac{2\sqrt{2}}{f_a m_a} \int_0^\pi \dd\theta \sqrt{\left[V(\theta) - V(0)\right]} \, , \label{eq:sigma_dw}
\end{align}
where $V$ is the axion potential. \cite{Huang:1985tt} used the full LO potential for axions, such that
\begin{equation}
    \mathcal{I} = \mathcal{I}(\zeta) = 4\, \sqrt{\frac{2}{\zeta}} \, \int_0^\pi \! \dd\alpha \; \sqrt{1 - \sqrt{1 - \zeta \sin^2(\alpha)}} \, ,
\end{equation}
where $\zeta \equiv 4z/(1+z)^2$ and $z \equiv m_u/m_d$ is again the ratio of the up and down quark masses. The numerical factor $\mathcal{I}(\zeta)$ takes values between $8$ and~$16 \times (2-\sqrt{2}) \approx 9.37$ for $\zeta \in [0,1]$, with $\mathcal{I}(0.87) \approx 8.9$ corresponding to the SM value. The result for the NLO potential is only $0.1\%$ larger \citep{1511.02867}.

However, the axion potential at temperatures $T \gtrsim T_\chi$ cannot be obtained from ChPT and we should, in line with \cref{eq:kge:potential}, use the simple cosine (dilute instanton gas) potential instead, for which
\begin{equation}
    \mathcal{I} = 2 \sqrt{2} \, \int_0^\pi \! \dd\theta \; \sqrt{1 - \cos(\theta)} = 8
\end{equation}
exactly and independently of the value of~$\zeta$.

Similar to the parameter~$\xi$ in \cref{eq:scaling_parameter} for strings, one can define the DW area parameter $\mathcal{A}$~\citep{1202.5851},
\begin{equation}
    \mathcal{A}(t) \equiv \rho_\text{dw}(t) \, t / \sigmaDW(t) \, ,
\end{equation}
which can be obtained from simulations and allows for a trivial computation of the axion energy density.
When the string-DW network starts decaying around time~$\tdecay$, the energy density of the network is given by 
\begin{equation}
    \rho_\text{s-dw}(\tdecay) = \rho_\text{s}(\tdecay) + \frac{\decay{\mathcal{A}} \, \sigmaDW(\tdecay)}{\tdecay} \, , \label{eq:energy_density_str_dw}
\end{equation}
where $\mathcal{A}_\text{decay} \equiv \mathcal{A}(\tdecay)$ and $\rho_\text{s}$ is given by \cref{eq:energy_density_strings}.

The decay of strings and DWs into axions ends at some time~$\tend$. However, it is not straightforward to determine -- or rather define -- precisely when this happens. For example, \cite{1202.5851} propose to define $\tend$ as the time when $\mathcal{A}(\tend) = 0.01$ i.e.\ from when on the domain wall area is less than one percent of the Hubble scale~[a 10\% criterion has also been examined in \cite{1412.0789}]. Typically, one finds that $\tend \approx \tdecay$~\citep{1202.5851,1412.0789}, such that one may simply take $R(\tdecay) \approx R(\tend)$ and $m_a(\tdecay) \approx m_a(\tend)$.

However, the subtle difference between the time scales is relevant for determining the average energy of axions. This is because the axion number density depends on how relativistic the average energy of axions at the time of emission, $\tilde{\epsilon}_\omega$, which may be defined as a multiple of the axion mass via $\bar{\omega}_a \equiv \tilde{\epsilon}_\omega \, m_a$ with a new parameter~$\tilde{\epsilon}_\omega$. Following \cite{1412.0789}, one may then write
\begin{align}
    n_a^\text{decay}(t_0) \approx \frac{\rho_\text{s-dw}(\tdecay)}{\bar{\omega}_a(\tend)} \left(\frac{R(\tdecay)}{R(t_0)}\right)^3 \approx \frac{\rho_\text{s-dw}(\tdecay)}{\tilde{\epsilon}_\omega \, m_a(\tdecay)} \left(\frac{R(\tdecay)}{R(t_0)}\right)^3 \, . \label{eq:energy_density_str_dw_decay}
\end{align}

This leaves $\decay{\mathcal{A}}$ and $\tilde{\epsilon}_\omega$ to be determined from simulations. The relic density from the string-domain wall decays today can thus be obtained from \cref{eq:energy_density_str_dw_decay}, which we show in \cref{fig:axion_energy_density}.

\subsection{Assumptions and limitations}\label{sec:limitations}
Let us comment on the underlying assumptions and limitations of this work.

While we focus on the well-defined statistical uncertainties, it is clear from the range of estimates of the DM QCD axion mass in the literature that systematic effects are the larger source of uncertainties.

However, the differences between the codes could in part be due to the different assumptions made and numerical schemes used for the simulations.
The procedures used for extracting the spectra and other information from the simulation data also come with their own uncertainties.
Obtaining estimates for these, using a principled approach, requires intimate understanding of the simulation codes.
We are thus not sure to what extent the estimates stated in the literature are indeed of that nature, given that they are often derived as an educated guess or not available at all.

Furthermore, it is often claimed that the extrapolation over many orders of magnitude introduces by far the largest source of systematic uncertainties.
While the fitting formulae used for the extrapolation might turn out to not be appropriate, in which case we cannot estimate the size of the systematic effects, there is no reason to treat this differently from any other fitting problem.
It should therefore be possible to determine the results under various hypothesis, compare the quality of the fit to the data, and estimate the size of the systematic uncertainty of the DM abundance today -- even though this is notoriously more involved than propagating statistical uncertainties. 

The only way to avoid models and fitting formulae is to perform simulations up to the physical scales, which does not seem feasible for the foreseeable future.
Nonetheless, it would be re-assuring to run a longer simulation, showing the scaling violation in $q$ observed in GHV-II over a longer period, and crossing explicitly (rather than by extrapolation) into the $q>1$ IR dominated regime.\footnote{Again noting that e.g.\ \cite{2108.05368} contradict the existence of a scaling violation in $q$ out to $\log \sim 9$ (although they support the existence of scaling violation in $\xi$).}

Differences between the simulations are further introduced by the fact that the derivation of the string and DW parameters cannot be done self-consistently since the computational cost of the simulations makes it difficult to perform the simulations for a number of QCD~axion parameters. For example, \cite{1202.5851} set $p = 6.68$ based on results from the interacting instanton liquid model~\citep{0910.1066}, but then choose $\chi_0^{1/4}, T_\chi \gtrsim \SI{5e15}{\GeV}$ for their simulations. This was done so that $\Tosc \sim T_\chi$ for their adopted value of $f_a = \SI{1.23e17}{\GeV}$, which simplifies the numerical simulations. These values are not even marginally consistent with realistic QCD axion parameters~(cf.\ Sec.~\ref{sec:paramsqcd}). It would hence be desirable to perform string and DW simulations for a more realistic, and consistent set of parameters for the string scaling and string-DW decay regimes as well as with different initial conditions if one cannot go deep into the scaling regime.

With that said, we decided to only consider the well-defined statistical uncertainties since these can be readily estimated.
While a detailed study of the different sources of systematic uncertainties -- along with comparison of the fitting formulae and physical models for the evolution and decay of topological defects -- would be desirable, this effort goes beyond the scope of the present work.

In terms of assumptions, we take the Universe to be radiation-dominated until we can apply \cref{eq:edescaling}. This assumption may be violated if the Universe becomes matter-dominated earlier than in standard cosmology [e.g.\ due to moduli, see \cite{0912.0015,Visinelli:2018wza}]. Furthermore, \cref{eq:edescaling} is only valid if entropy is (approximately) conserved in the later evolution. Significant entropy injection or other processes can spoil this assumption. 

It should also be noted that some authors have estimated the uncertainties on~$\gR$ and~$\gS$~\citep{1803.01038} and their derivatives, which are between 7--12\% in the relevant temperature range. There is no unique way to include such uncertainties, but we do not need to since we estimate their effect on on the (subdominant) realignment energy density to be only between 3--4\%. We also neglect the contribution of thermal axions to $\gR$ and $\gS$ when solving the axion field equation in \cref{eq:klein_gordon}. Even if axions would always fully contribute (i.e. be thermalized), the systematic increase of the effective degrees of freedom themselves would only be between 1--5\%, which is smaller than the 7--12\% uncertainty on $\gR$ and $\gS$ mentioned above, and thus also negligible.

Finally, we only consider axion models with DW number $\NDW = 1$. For axion models with $\NDW > 1$, such as some  \cite[e.g.][]{1610.07593,2003.01100,2107.12378} of the so-called KSVZ models~\citep{1979_kim_ksvz,1980_shifman_ksvz}, the definition of the decay constant has to be modified, $f_a \mapsto \NDW f_a$, leading to multiple minima in the axion potential and creating different kinds of stable DWs. As mentioned before, stable DWs can easily produce more than the observed amount of DM, such that additional mechanisms are required to make this scenario viable.

\section{Parameter estimates}\label{sec:params}
Apart from the values of $f_a$ and $\lr$, we can obtain estimates for the other parameters from experiment, theory, or simulations. Here we discuss how to estimate these nuisance parameters from the literature.

\subsection{QCD axion properties}\label{sec:paramsqcd}
\begin{figure}
    \centering
    \includegraphics[width=4.25in]{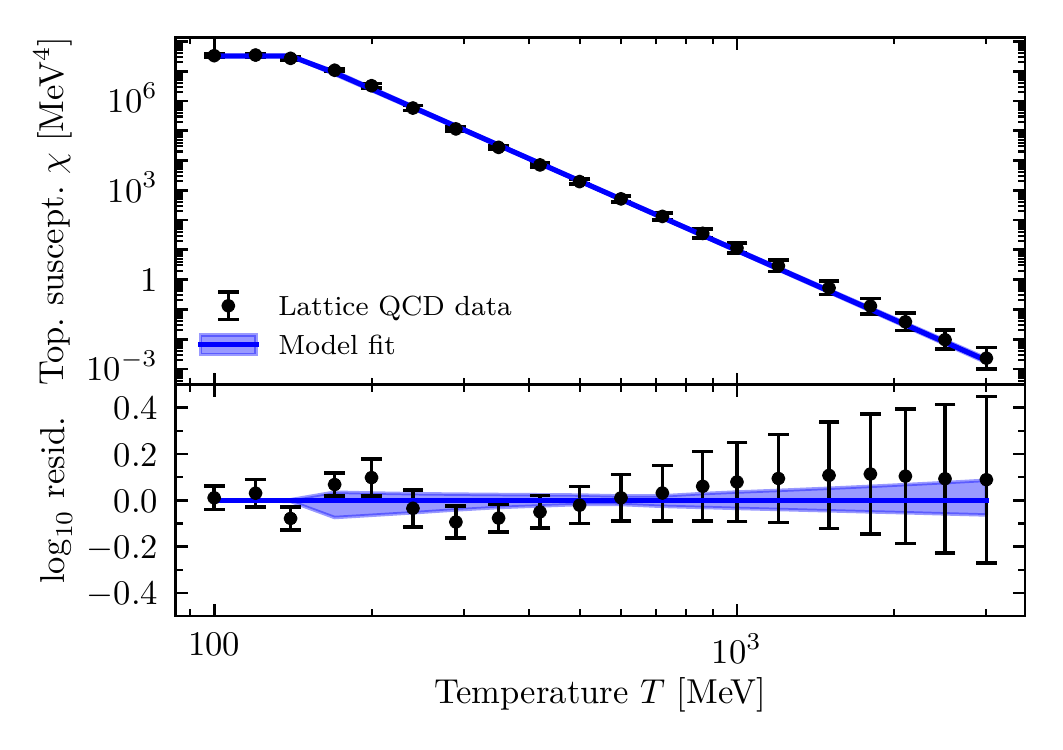}
    \caption{Fit to lattice QCD data. \textit{Top:} The fit of the power-law axion mass model~(blue line and shading) to the lattice QCD data~(black dots and error bars). \textit{Bottom:} Residuals (in log space) relative to the fitted model. Note that we only show the lattice QCD data, but simultaneously also fit $\chi_0$ from NNLO chiral perturbation theory.}
    \label{fig:qcd_axion_parameters}
\end{figure}
To estimate the QCD~axion parameters $\chi_0^{1/4}$, $T_\chi$, $p$, and $z$, we follow the strategy of \cite{1812.01008} and vary all quantities that enter the computation of Eq.~\eqref{eq:qcdaxion:zeromass:nnlo}, in particular~$z$, while fixing $f_\pi = \SI{92.3}{\MeV}$ and $m_\pi = m_{\pi^0} = \SI{134.98}{\MeV}$. This allows us to draw \num{e5}~samples for $\chi_0$ and $z$, which we then use to fit the remaining parameters~($T_\chi$ and $p$) to the latest results lattice~QCD results for temperatures~$T > 0$.\footnote{We use the data provided in Table~S7 of \cite{LatticQCD4Cosmo}.} Using a bootstrapping algorithm at the same time, i.e.\ generating data sets of the same size as the original data set but with (possibly multiple) random occurrences of the original data in each step, we obtain estimates for the parameters and uncertainties via the mean and covariance matrix from the sample of best-fitting points:
\begin{alignat}{3}
    \chi_0^{1/4} &= \SI{75.43 \pm 0.34}{\MeV} \, , \quad &&p &&= \num{7.75 \pm 0.11} \, , \nonumber\\
    T_\chi &= \SI{143.7 \pm 2.9}{\MeV} \, ,  \quad &&z &&= \num{0.472 \pm 0.011} \, ,
\end{alignat}
where we only quote the diagonal errors of the covariance matrix for simplicity, while we use the full covariance matrix in this work; the most sizable relative correlation coefficients~$\varrho$ are found between $T_\chi$ and $p$~($\varrho \approx 0.95$) and $p$ and $z$~($\varrho \approx 0.44$). We show the agreement with the high-temperature lattice QCD data in Fig.~\ref{fig:qcd_axion_parameters}.

\subsection{String spectrum: GHV-II}\label{sec:params:str:gorghetto}
GHV-II is our preferred work for estimating the parameters relevant for string spectrum, viz.\ $\xi$, $x_0$, and~$q$. Their results present some of the most realistic simulations to-date, with results presented in a way to allow us to infer the time dependence of the parameters through the instantaneous emission spectra $F[x,y]$, as defined in \cref{eq:def_F} and discussed in Sec.~\ref{sec:strings}.

First, consider the string scaling parameter~$\xi$. As mentioned before, there is evidence that~$\xi$ has an attractor solution such that, independent of the initial conditions, its value will asymptotically tend towards a function linear in~$\ell$. GHV demonstrate this by generating different initial conditions and computing $\xi$ as a function of time. In particular, GHV-II proposed to describe the resulting, initial deviations from the scaling behavior by using an ansatz of the form
\begin{equation}
    \xi = \xi_{-2} \, \ell^{-2} + \xi_{-1} \, \ell^{-1} + \xi_0 + \xi_1 \, \ell \, , \label{eq:xi_fit_form}
\end{equation}
where $\xi_{-2}$, $\xi_{-1}$, $\xi_0$, and $\xi_1$ are the fitting parameters, and where we again defined $\ell \equiv \log(m_r/H)$. Following GHV-II, we only consider data with $\ell \geq 4$~(except for one data set where $\ell \geq 5.5$) to reduce the impact of the initial conditions and we take $\xi_1$ as a universal parameter for the fit. Unlike GHV-II, however, we assign one $\xi_0^{(d)}$ parameter to each data set~$d$ instead of using it as a universal parameter.
We then perform a bootstrapping approach, where we calculate the parameter $\tilde{\xi}_0 = \text{median}(\xi_0^{(d)})$ as a proxy for the ``global'' $\xi_0$ parameter. Doing so for the data displayed in Figure~1 of GHV-II, we estimate
\begin{align}
   \xi_0 = \num{-1.618 \pm 0.038} \, , \quad  \xi_1 = \num{0.2428 \pm 0.0025} \, ,
\end{align}
and find to be essentially perfectly anti-correlated ($\varrho \approx -1$).
The values of $\xi_{-2}$ and $\xi_{-1}$ are not relevant for determining the axion energy density since the asymptotic, linear behavior of~$\xi$ dominates in the physical regime $\ell \gg 10$. We also consider the constant term since, at $\ell = 70$, it adds a contribution to $\xi$ at the 9\% level.

This leaves the parameters $x_0$ and $q$ to be determined. To do so, we extend the \textit{ansatz} given in \cref{eq:instantaneous_emission_spectrum_form} to also include the regions of $x < x_0$ and $x > y$. This is because we need to identify the cutoff~$x_0$ in the context of the surrounding spectrum. We therefore use a generalized form of the spectrum for fitting, which is given by
\begin{equation}
    F[x,y] \propto \frac{\left(\frac{x}{x_0}\right)^{q'} \left[1 + \Theta(x - x'')\left(\left(\frac{x''}{x}\right)^{q'' - q} - 1\right)\right]}{\left(\frac{x}{x_0}\right)^{q' + 1} + 1} \, , \label{eq:ies_extended_form}
\end{equation}
where $\Theta(\cdot)$ is the Heaviside function and $x'' \sim y$, $q'$, and $q''$ are additional fitting parameters~\citep{1806.04677}. Since the parameters $q$ and $x_0$ are the only relevant parameters for computing axion number density in \cref{eq:gorghettos_number_density_formula}, the additional fitting parameters can be ignored. The extended form in \cref{eq:ies_extended_form} also includes information on lower and higher momenta and makes the procedure less sensitive to cuts on the simulation data.

In GHV-I, the authors speculated about a time dependence of~$q$, for which further evidence was found in GHV-II. The authors found that a linear model seems to provide a good fit to the data, such that
\begin{equation}
    \label{eq:q_scaling}
    q = q_0 + q_1 \, \ell \, ,
\end{equation}
where $q_0$ and $q_1$ are fitting constants.

\begin{figure*}
    \centering
    {
    \hfill
    \includegraphics[width=3in]{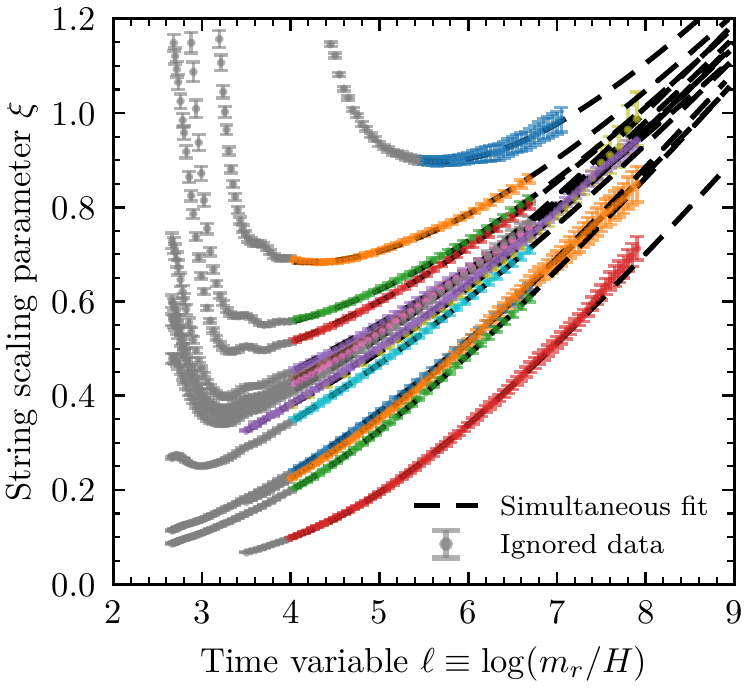}
    \hfill
    \includegraphics[width=3in]{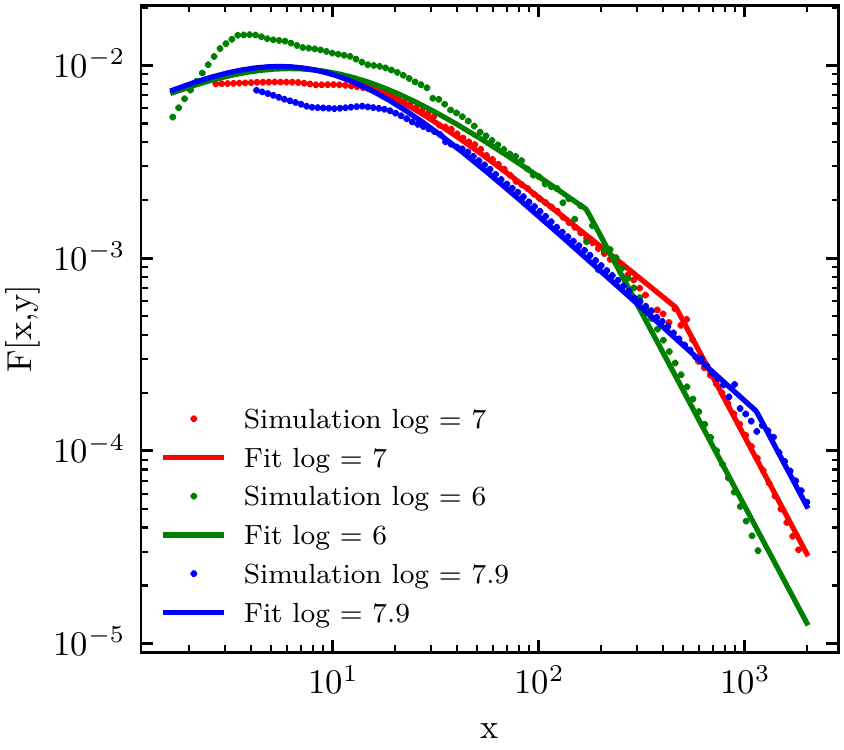}
    \hfill
    }
    \caption{Fitting the string scaling parameter~$\xi$ from Figure~1 and instantaneous emission spectra $F[x,y]$ from from Figure~14 of GHV-II.
    \textit{Left:} Simultaneous fit~(dashed lines) of~$\xi$ as a function of $\ell$, using \cref{eq:xi_fit_form} with $\xi_1$ being a universal parameter. The data points are shown in different colours, while gray data points are ignored.
    \emph{Right:} Simultaneous fit of spectra for $\ell = 6,7,7.9$~(green, red, and blue) as a function of the normalized momentum $x$. Data points are shown as dots, while solid lines indicate the fits.}
    \label{fig:new_gorghetto_fit}
\end{figure*}
We fit the spectra for physical strings from Figure~14 in GHV-II simultaneously, such that $x_0$, $q_0$, and $q_1$ are universal parameters and the additional parameters in \cref{eq:ies_extended_form} are specific for each spectrum. We apply smoothing to the spectra and interpolate them to obtain data at equally spaced $x$-values in log space. 
This is done to avoid over-weighting parts of the spectra compared to others.
The results are shown in \cref{fig:new_gorghetto_fit} and we estimate the fitting parameters to be
\begin{equation}
    q_0 = \num{0.154 \pm 0.057} \, , \quad q_1 = \num{0.1030 \pm 0.0071} \, , \quad x_0 = \num{8.22 \pm 0.83} \, ,
\end{equation}
where we only quote the diagonal errors of the covariance matrix while, in reality, we use full correlations in our analysis.

\subsection{String spectrum: HKSYY}\label{sec:params:str:hksyy}
While GHV-II is our preferred reference for the string contributions, the author do not consider the contribution from string-DW decay. For the latter, we use \cite{1202.5851,1412.0789}, referred to as HKSS and KSS hereafter. It would be desirable to also use the string parameters from the string-DW simulations to be consistent. However, HKSS point out that, unlike the DW parameters, the string parameters from their study are not expected to be reliable. Instead, one should use the results from their earlier studies~\citep{hep-ph/9811311,1012.5502}, of which we choose \cite{1012.5502}~[HKSYY hereafter; HKSSYY refers to \cite{1012.5502,1202.5851,1412.0789} collectively].

HKSYY find that the string scaling parameter~$\xi$ approaches a constant value, which is why we set $\xi_1 = 0$ and 
\begin{align}
    \xi_0 = \num{0.87 \pm 0.14} \, , \quad \xi_1 = 0 \, .
\end{align}
Note that this value is much smaller than the value found in GHV-II, whose results translate to $\xi \sim 15$ in the physical regime~($\ell \sim 70$).

While HKSYY do not use the same formalism as GHV, i.e.\ the instantaneous emission spectrum~$F[x,y]$. They do, however, show power spectra~$P$, defined via the number density as $n = \frac{1}{2\pi}\int \! \dd k \, P(k,t)/k$. Comparing this definition to \cref{eq:number_density_from_spectrum}, one cannot directly infer $F[x,y]$ since we require the ``free'' spectrum at different times, which is why HKSYY provide the difference spectrum
\begin{equation}
   \Delta P(k, t', t'') \equiv R^4(t'') \, P(k, t'') - R^4(t') \, P(k, t') \, , \label{eq:kawasaki_diff_spectrum}
\end{equation}
where $t' = 12.25\, t_\text{c}$, and $t'' = 25 \, t_\text{c}$, with $t_\text{c}$ being the time of PQ~symmetry breaking and where $P$ is only the ``free'' part of the spectrum. 
This is indeed proportional to a finite difference approximation of
the instantaneous spectrum $F[x,y]$.
However, since the separation between the two times $t'$ and $t''$ is relatively large ($\Delta \ell \sim 0.7$), we do not use this finite difference as an approximation to $F[x,y]$ but construct our fitting function in the following way:

First, we compute the total spectra $P(t')$ and $P(t'')$ from the instantaneous spectrum $F[x,y]$ by numerically evaluating the time integral in \cref{eq:energy_spectrum_rel}.
The emission rate $\Gamma(t)$ is then computed via numerical integration of the momentum integral of $\partial \Gamma / \partial k$, which is related to $F[x,y]$ via \cref{eq:def_F}.
We again assume $F[x,y]$ to have the form given in \cref{eq:ies_extended_form}.
This allows us to compute $\Delta P(k, t', t'')$ from a given set of parameters for $F[x,y]$ and, finally, a least-squares fit with bootstrapping gives
\begin{equation}
    q_0 = \num{1.44 \pm 0.16} \, , \quad q_1 = 0 \, , \quad x_0 = \num{2.2 \pm 1.2} \, ,
\end{equation}
where $q_1 = 0$ was set to zero in line with the findings of HKSYY, and $q_0$ and $x_0$ have a relative correlation coefficient of $\varrho \approx  0.75$.

\begin{figure}
    \centering
    \includegraphics[width=4in]{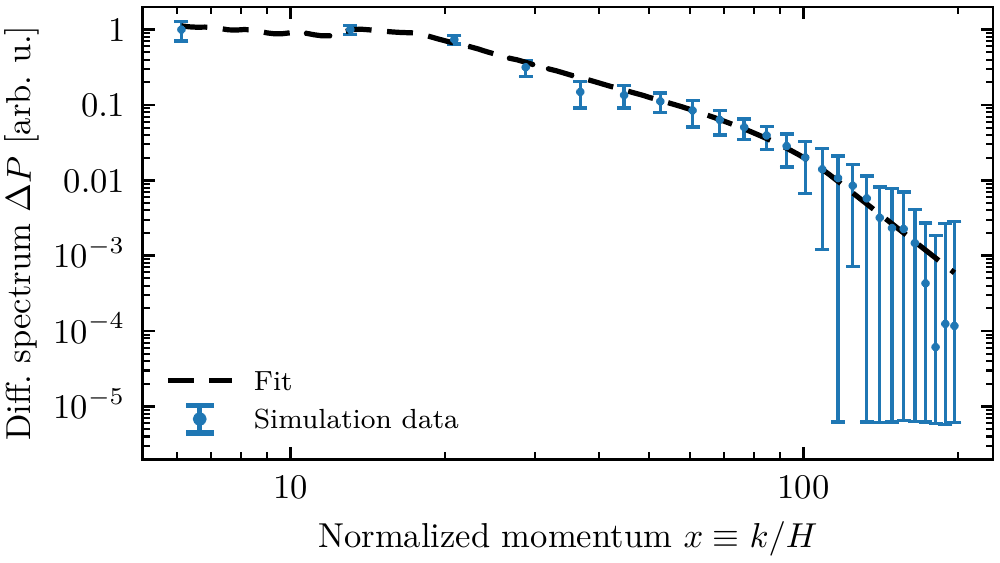}
    \caption{The differential energy spectrum $\Delta P$ between $t = 12.25 t_\text{c}$ and~$25t_\text{c}$ from Figure~7 in HKSYY as a function of normalized momentum~$x=k/H$ in arbitrary units. We show the data points with uncertainties~(blue) and the fitted spectrum~(dashed line).}
    \label{fig:kawasaki_spectrum_fit}
\end{figure}

\subsection{Nonlinear transient: GHV-II}
GHV-II estimate the numerical values constants for the nonlinear transient as $c_n = 1.35$, $c_m = 2.08$, and $c_V = 0.13$, and we implement them as fixed values without uncertainties. This clearly requires a justification.

By digitizing the data in Figure~3 of GHV-II and fitting the three quantities above, we find that the assumption of these being indeed constant to be justified. However, there is some residual dependence on the other model parameters, which we cannot fully incorporate as the the information is only provided for specific parameters. To estimate the size of the effect, we fit the curve for $x_0 = 10$~(which is consistent with the estimated~$x_0$) and $p = 8$~(which consistent with our estimate for this parameter) and find that $c_n = \num{1.48\pm 0.24}$, $c_m = \num{2.21 \pm 0.39}$, and $c_V = \num{0.114 \pm 0.039}$ from the fit. Using Eq.~(36) in GHV-II, we estimate that this introduces an error on~$\rhos$ of the order of~1.4\%. Had we done an inconsistent, simultaneous fit to the curves with $x_0 = 5$ and $x_0 = 10$~(which bracket the estimated~$x_0$), the error would be around~$2.2\%$.

In either case, this error is somewhat smaller than the estimated size of the other statistical effects, which we estimate to be around~8\% for the relevant values of~$f_a$.
In fact, since we cannot determine these parameters as a function of the other model parameters due to fitting to two inconsistent values for~$x_0$ simultaneously, the error estimates are not fully statistical but also of systematic nature.
The actual effect would therefore be smaller.
In conclusion, we think it is reasonable to use the values provided in the reference and ignore the associated uncertainties.

\subsection{String and domain wall decays: (H)KSS}\label{sec:paramsdws}
The parameters for the string and DW decay can be inferred from the string parameters estimated in the previous parts on the one hand, and from the works of HKSS and the updates from KSS on the on the other hand. In particular, we only require estimates for $\mathcal{A}_\text{decay}$ and $\tilde{\epsilon}_\omega$.

The authors perform multiple simulation for each of the chosen values of their $\kappa$ parameter, which effectively determines the value of $\chi_0^{1/4} \propto \kappa \, f_a$ w.r.t.\ their adopted $f_a = \SI{1.23e17}{\GeV}$. While none of their choices are for realistic for QCD~axions, we determine the parameters for $\kappa = 0.275$~(corresponding to $\chi_0^{1/4} \approx \SI{9e16}{\GeV}$) since this the lowest value of $\kappa$ for which we can self-consistently determine the parameters. We estimate directly from KSS a value for $\tilde{\epsilon}_\omega$, while we need to average the results for $\kappa = 0.25$ and $\kappa = 0.3$ from HKSS to estimate $\mathcal{A}_\text{decay}$:
\begin{align}
    \mathcal{A}_\text{decay} = 0.465 \pm 0.043\, , \quad \tilde{\epsilon}_\omega = 3.23 \pm 0.18 \, . \label{eq:paramsdws}
\end{align}

Three comments on these estimates are in order.
The first is that we do not include a possible correlation between the uncertainties as we cannot directly obtain it from the figures. However, by comparing the values for the data points for $\kappa \in \{0.3, 0.35, 0.4\}$, there seems to be a hint of a correlation coefficient of about $\varrho \sim -0.5$ between $\tilde{\epsilon}_\omega$ and $\mathcal{A}_\text{decay}$. Unfortunately, three data points do not allow us to establish reliable estimate, even though we expect that a correlation exists.

Further note that (H)KSS adopt a value of $\mathcal{I} = 9.23$ in \cref{eq:sigma_dw}. As a consequence, the value for $\mathcal{A}_\text{decay}$ needs to be rescaled by a factor of $9.23/8 \approx 1.15$ to be self-consistent with the calculated energy density.

Finally, (H)KSS guess an uncertainty of $\mathcal{A}_\text{decay} = 0.50 \pm 0.25$ for their estimates of the energy density in axions. This guess seems to include the sizeable systematics that exist, and which cannot be estimated well. In the spirit of this work, we only include the statistical uncertainties as present in the data and quoted in \cref{eq:paramsdws}.

\section{Constraints and likelihood}\label{sec:constraints}
The relevant constraints for this work come from the measurement of relic DM density, $\Omega_\text{c}h^2$, and the effective number of neutrino species, $\Neff$, defined via
\begin{equation}
    \rho_\text{rad} = \frac{\pi^2}{15} \left[1 + \frac{7}{8} \left(\frac{4}{11}\right)^\frac{4}{3} \Neff \right] \, T^4 \, ,
\end{equation}
where $\rho_\text{rad}$ is the energy density in relativistic degrees of freedom. The SM prediction of $\Neff$ today, $\Neff^{\text{\tiny SM}} = 3.045$~\citep{1606.06986}, then allows us to define the difference
\begin{equation}
    \Delta\Neff = \Neff - 3.045 \, .
\end{equation}

For a new particle beyond the SM that decouples from the thermal bath at a temperature of~$\Tdec$, $\Delta\Neff$ is given by
\begin{equation}
    \Delta\Neff = 0.027\,\left(\frac{106.76}{\gS(\Tdec)} \right)^{4/3} \, . \label{eq:axion_Delta_N_eff}
\end{equation}

\begin{figure}
    \centering
    \includegraphics[width=4in]{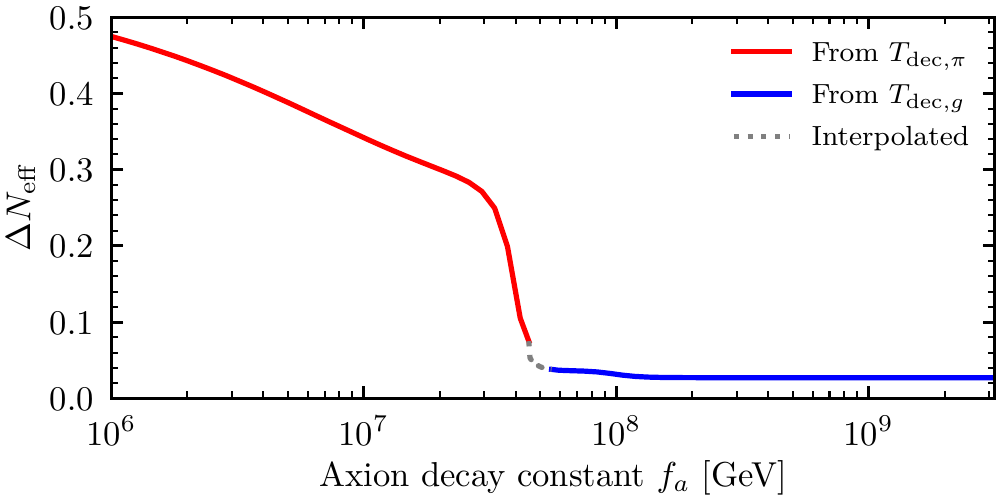}
    \caption{$\dNeff$ from the thermal production of axions as a function of the decay constant~$f_a$. We highlight where the decoupling temperature is determined by axion-pion~(red) or axion-gluon~(blue) interactions. Around the QCD cross-over, we interpolate these two regimes~(dotted grey line).\label{fig:Delta_N_eff_plot}\vspace{1.5em}}
\end{figure}

The main processes for decoupling of the QCD~axion model are, at low temperatures, axion-pion and, at high temperatures, the axion-gluon interactions. The decoupling temperature from gluons is implicitly given by
\begin{equation}
    \Tdecg = 2.50 \,  \frac{\sqrt{\gR(\Tdecg)}}{\alphaS^3(\Tdecg)} \frac{f_a^2}{\mpl} \, , \label{eq:T_d_gluon}
\end{equation}
where $\alphaS$ is the strong fine structure constant~\citep{2003.01100}.
We compute $\alphaS$ using the standard one-loop extrapolation from the $Z$~boson mass, as discussed in \citet[Sec.~9]{PDG} but ignoring threshold effects from the quark masses.\footnote{We thank \citet{2108.04259,2108.05371} for pointing out that more accurate results for the decoupling temperature can be obtained by solving the full Boltzmann equation instead, and that \citet{1310.6982} have already estimated $\Tdecg$ more accurately than the dimensional analysis estimate of \cref{eq:T_d_gluon}. This affects our estimates for $\Tdec$ and, to a much lesser degree, $\dNeff$. However, at the current level of sensitivity of cosmological surveys, $\Tdecg$ is not needed for estimating the upper limit of the axion mass window, cf.\ \cref{fig:Delta_N_eff_plot}.}
We use results of the OPAL collaboration at LEP, namely $M_Z = \SIsmp{91.187(7)}{\GeV}$~\citep{OPAL_mZ} and $\alphaS(m_Z^2) = 0.1189(43)$~\citep{OPAL_alphaS} as inputs for the calculation.
Using these results is preferable since the measurements from lepton colliders do not depend on complicated parton distribution functions.\footnote{We thank Enrico Bothmann for pointing this out.}
The typical uncertainty on~$\alphaS$ due to the experimental errors on~$M_Z$ is about~2\%. This is irrelevant for large~$f_a$ as the resulting error on~$\Tdecg$ has no influence due to constant~$\gS(\Tdecg)$. However, for lowest allowed values of $f_a \sim \SI{e8}{\GeV}$, we find an error of around~3\% on $\Delta\Neff$ (compared to an uncertainty of 6\% on the measured value of $\Neff$).

The decoupling temperature from pions is implicitly given by~\citep{2003.01100}
\begin{align}
    \label{eq:T_d_pion} 
    H(\Tdecpi) = 0.215 \, \gapi^2 \frac{\Tdecpi^5}{f_\pi^2 m_\pi^2} \, G\left( \frac{m_\pi}{\Tdecpi} \right) \, ,
\end{align}
where $f_\pi = \SI{92.3}{\MeV}$ and $m_\pi = m_{\pi^0} = \SI{134.98}{\MeV}$ are again the pion decay constant and mass, respectively. The function $G$~encodes the suppression of the interaction at higher pion masses, with $G(x) \simeq 1$ when $x \ll 1$, and can be computed numerically~\citep{hep-ph/0504059}.

Note that Eqs.~\eqref{eq:T_d_gluon} and~\eqref{eq:T_d_pion} are \emph{not} valid for temperatures around the QCD~crossover, $T \sim T_\text{QCD,c}$, as neither ChPT nor gluons are good descriptions in this regime. To provide an approximation, we interpolate the decoupling temperature linearly in~$\log_{10}(f_a/\si{\GeV})$. The resulting prediction for $\dNeff$, highlighting the region where we had to interpolate the resulting decoupling temperature, can be found in \cref{fig:Delta_N_eff_plot}. However, we also highlight a recent work that points out that the ChPT approach for calculating $\dNeff$ from axion-pion interactions needs to be revised even for $f_a \gtrsim \SI{5e6}{\GeV}$~\citep{2101.10330}. The assumptions made about the range of validity of ChPT are hence potentially somewhat optimistic. There are currently efforts underway to improve the estimate for $\Tdec$ in this region of parameter space~\citep{2108.04259,2108.05371}.

Further note that decoupling from the SM bath of course requires that the new degrees of freedom have been in thermal equilibrium before that point. In particular, the reheating temperature at the end of inflation should be higher than the expected decoupling temperature, $\Treh > \Tdec$. Requiring that this be the case for the axion-gluon interactions~(relevant at higher temperatures), one finds that~\cite[e.g.][]{1604.08614}
\begin{equation}
    f_a < \SI{2e11}{\GeV} \, \left(\frac{\alphaS(\Tdecg)}{0.03}\right) \left(\frac{\Treh}{\SI{e10}{\GeV}}\right)^\frac{1}{2} \, .
\end{equation}
This means that it is indeed possible that the constraints from $\dNeff$ do not apply when both $f_a$ and $\Treh$ are sufficiently low. However, we do not consider a model or constraints for inflation, and may therefore always apply the $\dNeff$ constraints but keeping in mind that this implicitly assumes a sufficiently high value of $\Treh$.

Given the prediction for $\Omega_\text{c} h^2$, which is calculated as the sum of the realignment contribution and all topological defect contributions under consideration in a setup, and $\Neff$ from \cref{eq:axion_Delta_N_eff}, we could perform a full cosmological analysis as e.g.\ conducted by the \textit{Planck} Collaboration. However, to simplify our setup and to allow us to easily relax the requirement for axions to match the dark matter density, we re-interpret the posterior from the analysis of \cite{1807.06209} as a likelihood,
\begin{equation}
    \log(L) = -\frac{1}{2} \, \vc{\mu}_\text{CMB} \, \Sigma_\text{CMB}^{-1} \, \vc{\mu}_\text{CMB}^\mathrm{T} \, ,
\end{equation}
where $\vc{\mu}_\text{CMB} \equiv (\Omega_a h^2 - \hat{\Omega}_\text{c} h^2,\, \Neff - \hat{N}_\text{eff})$ when QCD axions are all of the dark matter and $\vc{\mu}_\text{CMB} \equiv (\max [0,\Omega_a h^2 - \hat{\Omega}_\text{c} h^2],\, N_\text{eff} - \Neff)$ when we allow them to constitute only a fraction of the dark matter. Of course, this is not fully consistent with either Bayesian or frequentist philosophy but arguably acceptable due to the fairly well-constrained nature of these parameters. 
Using the latest \textit{Planck} data,\footnote{We use the \textit{Planck}~2018 chains~(\texttt{base\_\allowbreak{}nnu\_\allowbreak{}plikHM\_\allowbreak{}TTTEEE\_\allowbreak{}lowl\_\allowbreak{}lowE\_\allowbreak{}BAO\_\allowbreak{}post\_\allowbreak{}lensing}) from the \textit{Planck} Legacy Archive, available at \url{https://www.cosmos.esa.int/web/planck/pla}. We make use of the \texttt{cosmomc}~\citep{astro-ph/0205436,1304.4473} and \texttt{getdist}~\citep{1910.13970} packages to extract the posterior and to estimate the covariance matrix.} we find:
\begin{equation}
    \hat{\Omega}_\text{c} h^2 = 0.118 \, , \quad \hat{N}_\text{eff} = 2.985 \, , \quad
    \hat{\Sigma}_\text{CMB} = \twocov{\num{8.22e-6}}{\num{4.68e-4}}{\num{2.98e-2}} \label{eq:loglike:cov} \, .
\end{equation}

\begin{figure}
    \centering
    \includegraphics[width=4in]{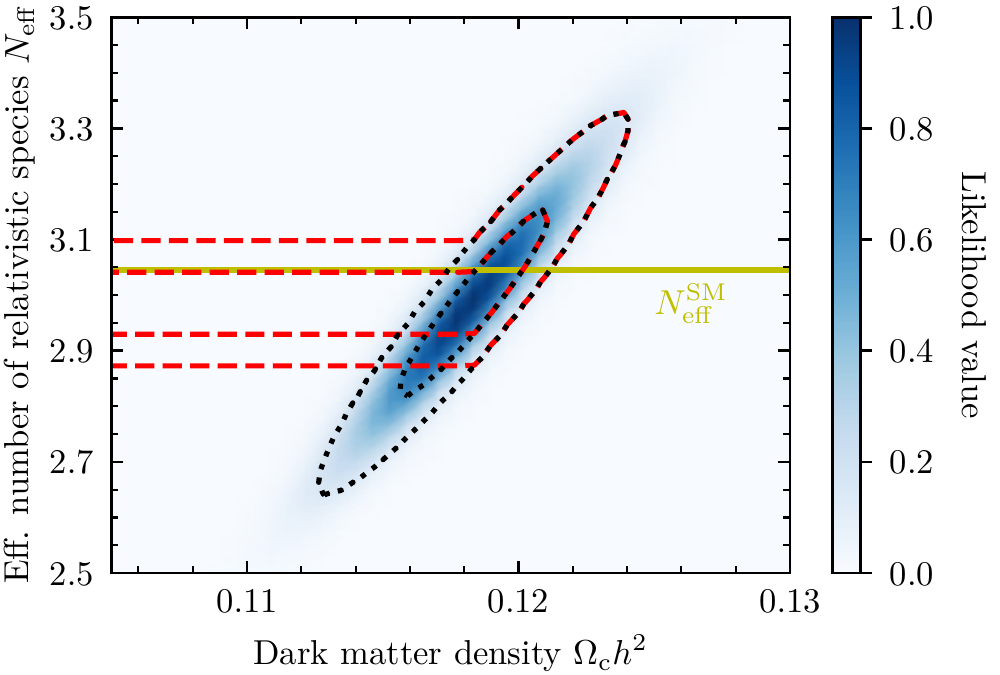}
    \caption{The joint likelihood~(blue color map) for the dark matter density, $\Omega_\text{c} h^2$, and the effective number of relativistic species, $\Neff$, constructed from the \textit{Planck} data. Dotted black~(dashed red) lines indicate the regions at 68\%/95\%~CL when axions are all of (allowed to to only be a fraction of) the dark matter density. A yellow line indicates the Standard Model value of~$\Neff$.\label{fig:planck_constraints}\vspace{1em}}
\end{figure}

The correlation between $\hat{\Omega}_\text{c} h^2$ and $\Neff$ in Eq.~\eqref{eq:loglike:cov} also visible in Fig.~\ref{fig:planck_constraints}, where show the two choices of likelihood as used in our analysis.

\section{Results}\label{sec:results}
We use the ensemble MCMC algorithm \texttt{emcee}~\citep{emcee} to sample our 12-\ or 13-dimensional parameter space, depending on whether we use the results of GHV or HKSSYY. The latter has three more parameters related to string-DW decay but at the same time two parameters less for the string scaling.

The priors for the QCD, string and DW nuisance parameters directly derive from our fits to the data in Secs.~\ref{sec:paramsqcd}--\ref{sec:paramsdws}. In agreement with GHV, we set $\lambda_r = 1/2$ and use a prior that is uniform in $\log_{10}(f_a/\si{\GeV}) \sim \mathcal{U}(7,16)$. The log-uniform prior encodes our ignorance to the scale of new physics, i.e.\ its order of magnitude, while the range approximately encompasses the possible range constrained by cosmological probes. Since the \textit{Planck} data is fairly restrictive, the results for our credible regions for the axion window will only very mildly depend on the exact location of the prior bounds; this is because the low likelihood outside of the axion window essentially results in zero posterior weight regardless of the prior.

While the string and DW parameters are potentially afflicted with large systematic uncertainties related to the validity of the underlying models and extrapolations, the appeal of this framework is that -- except for the choice of prior in~$f_a$ and~$\lr$ -- the prediction of the QCD~axion mass window is entirely based on physical information. Furthermore, since the physical priors constrain the model parameters rather well and independently of cosmological data, the only non-trivial posterior distributions are the ones on the axion decay constant~$f_a$ or, equivalently, on the axions mass~$m_{a,0} = \sqrt{\chi_0}/f_a$.

\begin{figure}
    \centering
    {
    \hfill
    \includegraphics[width=3in]{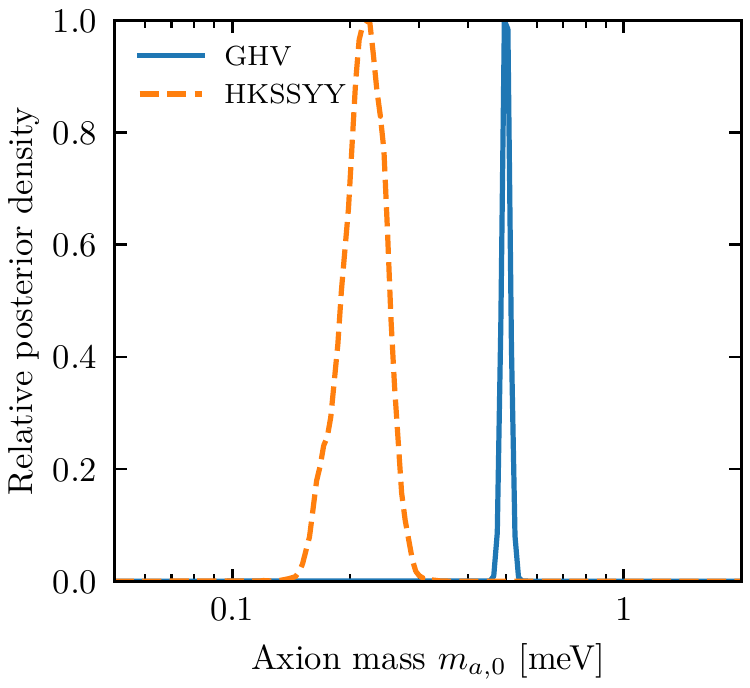}
    \hfill
    \includegraphics[width=3in]{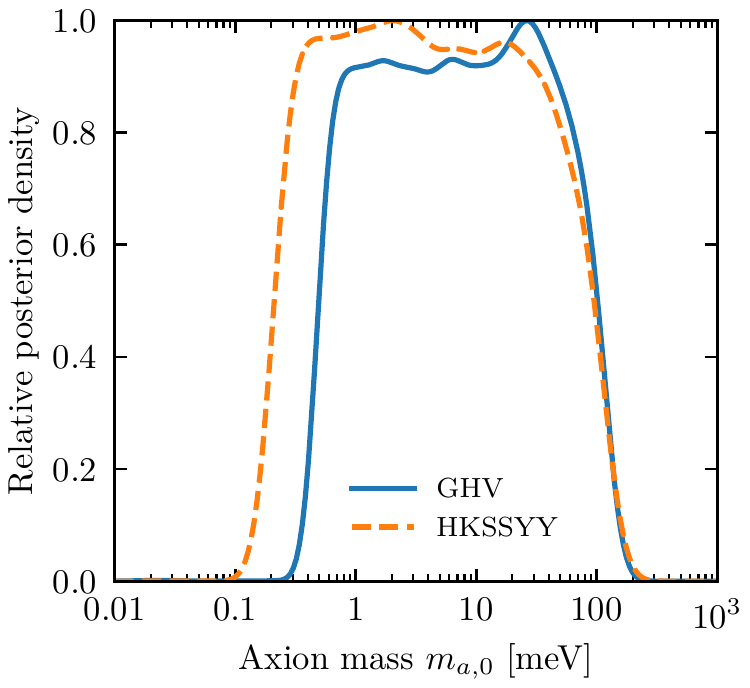}
    \hfill
    }
    \caption{Relative marginal posterior densities for the QCD axion mass when axions make up all of the dark matter in the Universe~(\textit{left}) and when axions are allowed to only make up a fraction of the total dark matter~(\textit{right}). We show the results for the GHV~(blue) and HKSSYY~(dashed orange) simulations, where the KDE of the posterior density in $\log_{10}(m_a/\si{\meV})$ is scaled to their respective maximum values.}
    \label{fig:mass_posterior}
\end{figure}
In the left panel of \cref{fig:mass_posterior}, we show the relative marginal posteriors, i.e.\ normalized to their maximum values, of the DM QCD~axion mass. Both distributions point to the mass range between \SIrange[range-phrase=~\text{and}~]{0.1}{1}{\meV} and are hence only constrained by the measured value of~$\Omega_\text{c}h^2$ and not by~$\Neff$. That being said, the predicted value of $\Delta\Neff \approx 0.027$ in both cases is only slightly worse than the SM prediction, when compared to the measured uncertainty on~$\Neff$ of about~0.173. The predicted value is also close to the sensitivity of future CMB missions \cite[e.g.][]{1604.08614,1610.02743}.

Using the GHV data, we find -- in agreement with that work -- a median QCD axion mass of \SI{0.50}{\meV}, while the 95\% credible interval at highest posterior density~(HPD) is $[0.48,0.52]\,\si{\meV}$. For HKSSYY the median and the interval are \SI{0.22}{\meV} and $[0.16,0.27]\,\si{\meV}$, respectively.

Na\"ively one would expect that the DM QCD axion mass should be higher for HKSSYY since we add the string-DW decay contribution. However, we remind the reader that the scaling violation is absent in the HKSSYY simulation, leading to a noticeably reduced contribution from the string scaling component. In this context, we also point out that the HKSSYY results imply that the number of axions from the decays is a factor of about 1.7 higher than that from string scaling~(or even more similar in magnitude when nonlinear transient effects are ignored). When adding a string-DW decay component to GHV, the estimated axion mass might therefore be even larger.

Finally, the left panel of \cref{fig:mass_posterior} also illustrates how the systematic uncertainties attached to the topological defect computations~(in this case different simulation codes and strategies) exceed the smaller statistical uncertainties. When other systematic effects were to be taken into account, we would thus expect them to dominate the total error budget.

In the next step we drop the requirement that QCD axions make up all of the DM in the Universe. In this case higher axion masses are possible, but only to the extent allowed by the constraint on~$\Neff$. Together with the dark matter constraints, the two fairly robust constraints from cosmology alone are sufficient to delimit the axion mass window in the post-inflationary PQ~breaking scenario, as shown in the right panel of \cref{fig:mass_posterior}. The 95\% HPD credible region encloses axion masses between $\SI{0.49}{\meV} < m_a < \SI{84}{\meV}$ for GHV and $\SI{0.23}{\meV} < m_a < \SI{82}{\meV}$ for HKSSYY.

Given that QCD axion DM is completely subdominant for axion masses $\gtrsim \SI{100}{\meV}$ where the $\dNeff$ constraints are relevant, the upper ends of the posteriors in the right panel \cref{fig:mass_posterior} overlap as the differences in the GHV and HKSSYY analyses are not relevant there.

\section{Conclusions}\label{sec:conclusions}
While some of the recent simulations suggest that the QCD axion energy density is dominated by topological defect scaling and decay, the debate surrounding this topic is not yet settled. Despite the differences in simulation results, one can use a common, parametric framework to infer the relevant model parameters as well as the associated (correlated) statistical uncertainties.

Cosmological probes then provide robust and sufficient constraints to define the QCD~axion mass window -- whether QCD~axions are a part or all of the dark matter in the Universe. Possible extension of this work could include constraints from e.g.\ astrophysics such as axion emission from supernova SN1987A due to axion-nucleon interactions~\citep{1906.11844}. More generally, the parametric approach taken here may also be useful for global fits of axion models \cite[e.g.][]{1708.02111,1809.06382,1810.07192} in the post-inflationary Peccei--Quinn symmetry breaking scenario, which might include indirect detection limits related to axion miniclusters \cite[e.g][]{Fairbairn:2017dmf,Edwards:2020afl}.

The main outcome of our analysis is that we explicitly demonstrate that statistical uncertainties from the available simulations and particle physics data alone are rather small, making this scenario in principle very predictive. For example, the GHV string-only simulations give a rather narrow range for the QCD axion mass of a few percent in case axions are all of the dark matter in the Universe. The upper end of the axion mass window comes from hot dark matter bounds and is independent of the topological defect calculation. This is very encouraging and provides further motivation for a better and more quantitative understanding of the systematic sources.

Following our discussion about the various systematic uncertainties in Sec.~\ref{sec:limitations}, the results of such a more detailed study of systematics could be incorporated into the framework used here in the future. It would further be useful to perform domain wall decay simulations with more physical parameter choices, which comes at the price of a reduced range of the simulation. Furthermore, the results of different groups could be compared by running string simulations with various algorithms for longer times under controlled conditions, similar to e.g.\ the AGORA project for galaxy formation simulations~\citep{1308.2669}.\footnote{Information about AGORA is available at \url{https://sites.google.com/site/santacruzcomparisonproject/}.} This might help to resolve some of the differences between the codes and provide further evidence for the (non-)existence of the scaling violation in~$\xi$, as well as explicitly demonstrate if $q > 1$ towards the end of the simulation or not. In this sense, our study is a first step in this direction of a more detailed comparison of different simulations, which seems necessary in light of the ongoing cycle of contradictory findings.

As for the upper end of the axion mass window, $\Neff$ has the potential to severely constrain the axion mass if the axion prediction for $\Neff$ can be made more robust~\citep{2101.10330,2108.04259,2108.05371} and if future CMB surveys will improve the sensitivity to this observable~\citep{1610.02743}. On the one hand, this requires an improved understanding of the interaction rates during the QCD phase transition, as discussed before. On the other hand, it may allow us to probe the Peccei--Quinn phase transition (via the axions from strings prediction) and the inflationary reheating temperature (via $\Neff$), offering a new window onto the early Universe~\citep{1604.08614}.

A more accurate determination of the lower end of axion mass window might also have important experimental consequences. If axions from topological defects dominate, the lower limit on the QCD axion mass is higher than what would be expected from the realignment contribution alone. If future studies confirm this picture, the case for the recent expansion of experiments such as e.g.\ the TOORAD proposal~\citep{1807.08810,2102.05366}, BRASS~\citep{1212.2970},\footnote{Information about BRASS is available at \url{https://www.physik.uni-hamburg.de/iexp/gruppe-horns/forschung/brass.html}.} or BREAD~\citep{2111.12103} that explore the meV~region of parameter space will be strengthened. On the other hand, a lower limit in the \SI{0.1}{\meV} range favors experiments such as MADMAX~\citep{Caldwell:2016dcw} or ALPHA~\citep{Lawson:2019brd}.

\section*{Acknowledgments}
We thank Marco Gorghetto and \cite{2108.04259,2108.05371,2109.09679} for helpful discussions about their works. SH and DJEM were supported by the Alexander von Humboldt Foundation and the German Federal Ministry of Education and Research. DJEM is now supported by the UK STFC on an Ernest Rutherford Fellowship. We acknowledge use of the Scientific Computing Cluster at GWDG, the joint data centre of Max Planck Society for the Advancement of Science~(MPG) and the University of {G\"ottingen} as well as computing resources of the North-German Supercomputing Alliance~(HLRN). We acknowledge use of the Python packages \texttt{emcee}~\citep{emcee}, \texttt{matplotlib}~\citep{matplotlib_1}, \texttt{mpi4py}~\citep{mpi4py_1,mpi4py_2,mpi4py_3}, \texttt{numpy}~\citep{numpy_1}, \texttt{pymc3}~\citep{pymc3}, \texttt{schwimmbad}, and \texttt{scipy}~\citep{scipy_1} as well as the \texttt{WebPlotDigitizer} tool~\citep{Rohatgi2020}.

\appendix

\section{Interpolation scheme for the number of effective relativistic degrees of freedom}\label{sec:g_star_appendix}
For calculating the axion energy density from the realignment mechanism~(see Appendix~\ref{sec:klein_gordon_appendix}), we need the values and derivatives of the effective relativistic degrees of freedom w.r.t.\ the energy density~$\gR$ and entropy~$\gS$.

\begin{figure}
    \centering
    \includegraphics[width=4.25in]{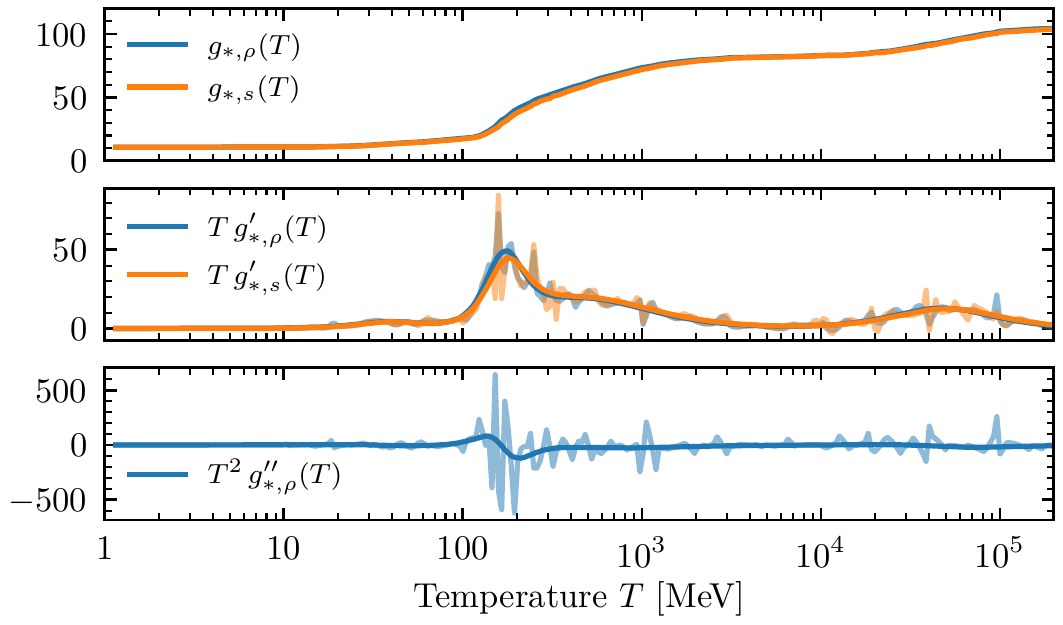}
    \caption{The effective relativistic degrees of freedom of energy and entropy density as functions of~$T$, as well as their derivatives. Note that the latter were made dimensionless by multiplying with appropriate powers of~$T$. The analytical derivatives from the fitting formula we used are shown as solid lines, while the derivatives directly obtained from the interpolating polynomials are shown as transparent solid lines.\label{fig:g_star_plot}}
\end{figure}
We use the values computed in \cite{LatticQCD4Cosmo} and then interpolate them in $\tau \equiv \log_{10} (T/\si{\MeV})$.
Since the given values are log-spaced, and separated by quite large steps, the derivatives cannot reliably be obtained from the interpolation, as can be seen in \cref{fig:g_star_plot}.
Instead, we fit the result using the ansatz
\begin{align}
    g_{*, i}(\tau) = \exp \left[ a^i + \sum_{j = 1}^5 b^i_j \left( 1 + \tanh \left( \frac{\tau - c^i_j}{d^i_j} \right)  \right) \right] \, ,
\end{align}
for $i=s,\rho$, which has previously been proposed by \citet{0910.1066}.
This allows us to evaluate the analytical derivatives of $\gR$ and $\gS$, which we then tabulate and interpolate  on a fine grid of $\tau$~values for better computational performance.

The numerical values of $\gR$ and $\gS$ and their relevant derivatives are shown in \cref{fig:g_star_plot}, together with the derivatives one would obtain from spline interpolations the raw data~(using default \texttt{scipy} interpolation routines).
It becomes clear that the derivatives from the interpolating polynomials are not reliable due to the rather large oscillating deviations seen in \cref{fig:g_star_plot}. This is in particular problematic for the second derivative of $\gR$ as e.g.\ a more reliable linear interpolation cannot be used here. While the tabulated data for $\gR$ and $\gS$ was rather sparse, we anticipate that this problem might even occur for more densely tabulated data.

\section{Solving the Klein--Gordon equation}
\label{sec:klein_gordon_appendix}
Let us discuss the numerical solution of the Klein--Gordon equation~\eqref{eq:klein_gordon}
\begin{align}
    \ddot{\phi} + 3 H \dot{\phi} + V'(\phi) = 0 \, ,
\end{align}
where we can ignore the spatial gradient term.
First, we re-scale the axion field~$\phi$ to the misalignment angle $\theta \equiv \phi / f_a$.
Since the axion potential, and hence the axion mass, is temperature-dependent, we need to track the temperature evolution as a function of time.
Alternatively, we may simply change variables from physical time~$t$ to the temperature of the photon bath~$T$, and obtain~[see also e.g.\ \cite{LatticQCD4Cosmo}]
\begin{align}
    \label{eq:klein_gordon_temperature}
    \frac{\dd^2 \theta}{\dd T^2} +
                 \left(
                 3H \frac{\dd t}{\dd T}
                 - \frac{\dd^2 t}{\dd T^2} \left/ \frac{\dd t}{\dd T} \right.
                 \right) \frac{\dd \theta}{\dd T} +
                 V'(\theta) \left( \frac{\dd t}{\dd T} \right)^2 = 0 \, ,
\end{align}
where
\begin{align}
    \label{eq:dtdT}
       \frac{\dd t}{\dd T} = - \mpl \sqrt{\frac{45}{8\pi^2}} \frac{T \gR'(T) + 4 \gR(T)}{T^3 \gS(T) \sqrt{\gR(T)}} \, .
\end{align}
We solve \cref{eq:klein_gordon_temperature} numerically using the \texttt{vode} algorithm from \texttt{scipy}.
The integration starts at $T = 5 \Tosc$, as already suggested in \cite{LatticQCD4Cosmo}.
We continue integrating \cref{eq:klein_gordon_temperature} until the first sign change of $\theta$ i.e.\ the start of the oscillations.

To ensure that we obtain an accurate result, we average the ratio of number and entropy densities, $n(T) / s(T)$, over a fixed number of oscillations, $N_\text{osc} = 3$, using $\Delta T / N$ separated points with the \texttt{simps} method of \texttt{scipy}.
Since the frequency of the oscillations does not become constant in $T$ but only in $t$, we need to adapt the step size that we use for the integration to ensure that we always have enough points within the integration interval. For this we assume that $\gR$ and $\gS$ are constant and, using the Friedmann equation, we find that when we start the averaging at $T_1$, then the interval with $N_\text{osc}$ oscillations is given by
\begin{align}
    \Delta T = \left[ \frac{\Delta t}{C} + \frac{1}{T_1^2} \right]^{-1/2} - T_1 \quad \text{with} \quad C = \sqrt{\frac{1440}{\gR(T_1)}} \, \mpl
\end{align}
and, from the frequency of the axion field in the WKB approximation,
\begin{align}
    \Delta t = \frac{2\pi \, N_\text{osc}}{m_a(T_1)} \, .
\end{align}

The above procedure is repeated until the relative change in the obtained $n/s$ values between two consecutive intervals is below the required precision of $\epsilon = \num{e-4}$.

\begin{figure}
    \centering
    \includegraphics[width=4in]{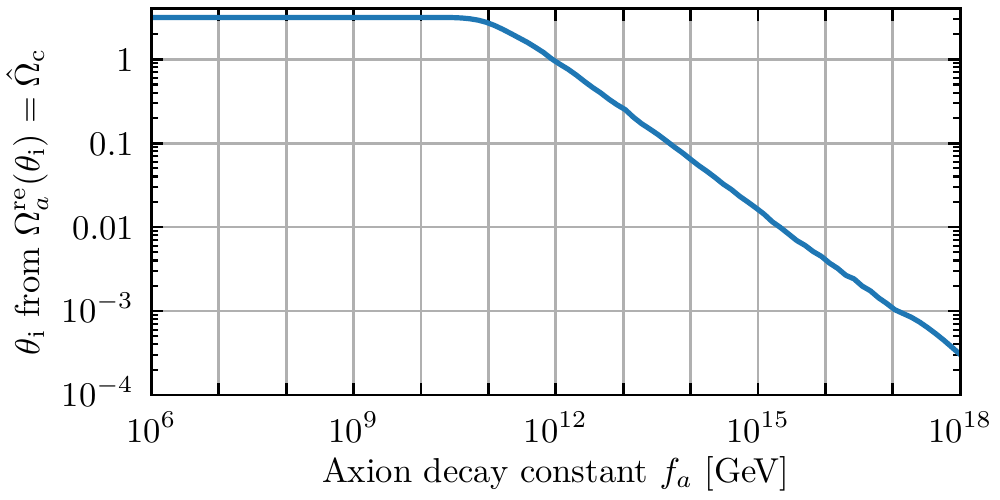}
    \caption{Values of the initial misalignment angle $\thetai$ for which the energy density from the realignment mechanism matches the dark matter density as a function of the axion decay constant $f_a$.}
    \label{fig:dm_axions_plot}
\end{figure}
To enable a comparison with our algorithm, we show the required value of the initial misalignment angle~$\thetai$ such that axions are all of the DM with $\Omega_a^\text{re} h^2 = \hat{\Omega}_\text{c} h^2 = 0.12$ in \cref{fig:dm_axions_plot}.

\vspace{2cm}

\bibliographystyle{mnras}
\bibliography{references}

\end{document}

%% file: macros.tex
\newcommand{\mpl}{\bar{m}_\text{Pl}}

\newcommand{\alphaS}{\alpha_\text{s}}
\newcommand{\ee}{\mathrm{e}}
\newcommand{\order}[1]{\mathcal{O}(#1)}

\newcommand{\thetai}{\theta_\text{i}}
\newcommand{\gR}{g_{\ast,\rho}}
\newcommand{\gS}{g_{\ast,s}}
\newcommand{\Treh}{T_\text{reh}}
\newcommand{\Tdec}{T_\text{dec}}
\newcommand{\Tdecg}{T_\text{dec,$g$}}
\newcommand{\Tdecpi}{T_\text{dec,$\pi$}}
\newcommand{\Neff}{\ensuremath{N_\text{eff}}}
\newcommand{\dNeff}{\ensuremath{\Delta\Neff}}

\newcommand{\NDW}{N_\text{DW}}
\newcommand{\lr}{\lambda_r}
\newcommand{\Capi}{C_{a\pi}}
\newcommand{\gapi}{g_{a\pi}}

\newcommand{\rhos}{\rho_a^\text{str}}
\newcommand{\sigmaDW}{\sigma_\text{dw}}
\newcommand{\rhoc}{\rho_\text{crit}}
\newcommand{\osc}[1]{#1_\text{osc}}
\newcommand{\Tosc}{\osc{T}}
\newcommand{\tosc}{\osc{t}}
\newcommand{\decay}[1]{#1_\text{decay}}

\newcommand{\tdecay}{\decay{t}}
\newcommand{\tend}{t_\text{end}}

\newcommand{\vc}[1]{\boldsymbol{#1}}

\newcommand{\twocov}[3]{\left(\begin{array}{S[table-format=-1.2e-1]S[table-format=-1.2e-1]} #1 & #2\\ #2 & #3\end{array}\right)}

\newcommand{\dd}{\mathrm{d}}
\newcommand{\del}{\partial}

\newcommand{\SIsmp}[2]{\SI[separate-uncertainty=false]{#1}{#2}}

\newcommand{\tnl}{t_\text{nl}}
\newcommand{\Tnl}{T_\text{nl}}

\newcommand{\aem}[1]{\href{mailto:#1}{#1}}

%% file: journal_macros.tex
\def\apj{ApJ}%
\def\apjl{ApJ}%
\def\apjs{ApJS}%
\def\ao{Appl.\ Opt.}%
\def\aap{A\&A}%
\def\jhep{JHEP}%
\def\jcap{JCAP}%
\def\prd{Phys.\ Rev.\ D}%
\def\prl{Phys.\ Rev.\ Lett.}%
\def\pasp{PASP}%
\def\nat{Nature}%
\def\physrep{Phys.\ Rep.}%